\documentclass[letterpaper]{article} 
\usepackage[submission]{aaai2026}  
\usepackage{times}  
\usepackage{helvet}  
\usepackage{courier}  
\usepackage[hyphens]{url}  
\usepackage{graphicx} 
\usepackage{natbib}  
\usepackage{caption} 
\usepackage{algorithm}

\usepackage{newfloat}
\usepackage{listings}
\DeclareCaptionStyle{ruled}{labelfont=normalfont,labelsep=colon,strut=off} 
\floatstyle{ruled}
\newfloat{listing}{tb}{lst}{}
\floatname{listing}{Listing}
\title{Gotta Hear Them All: Towards Sound Source Aware Audio Generation}
\author {
    Wei Guo\textsuperscript{\rm 1},
    Heng Wang\textsuperscript{\rm 1},
    Jianbo Ma\textsuperscript{\rm 2},
    Weidong Cai\textsuperscript{\rm 1}
}
\affiliations {
    \textsuperscript{\rm 1}University of Sydney \hspace{4pt} \textsuperscript{\rm 2}Dolby Laboratories\\ \vspace{2pt}
    wei.guo@sydney.edu.au \quad heng.wang@sydney.edu.au \quad jianbo.ma@dolby.com \quad tom.cai@sydney.edu.au
}

\usepackage{algpseudocode}

\usepackage{amsmath}

\usepackage[capitalize]{cleveref}
\crefname{section}{Sec.}{Secs.}
\Crefname{section}{Section}{Sections}
\Crefname{table}{Table}{Tables}
\crefname{table}{Tab.}{Tabs.}

\usepackage{booktabs}
\usepackage{multirow}
\usepackage{amssymb}
\usepackage{pifont}
\newcommand{\cmark}{\ding{51}}%
\newcommand{\xmark}{\ding{55}}%

\begin{document}

\maketitle

\begin{abstract}
Audio synthesis has broad applications in multimedia. Recent advancements have made it possible to generate relevant audios from inputs describing an audio scene, such as images or texts. However, the immersiveness and expressiveness of the generation are limited. One possible problem is that existing methods solely rely on the global scene and overlook details of local sounding objects (i.e., sound sources). To address this issue, we propose a Sound Source-Aware Audio (SS2A) generator. SS2A is able to locally perceive multimodal sound sources from a scene with visual detection and cross-modality translation. It then contrastively learns a Cross-Modal Sound Source (CMSS) Manifold to semantically disambiguate each source. Finally, we attentively mix their CMSS semantics into a rich audio representation, from which a pretrained audio generator outputs the sound. To model the CMSS manifold, we curate a novel single-sound-source visual-audio dataset VGGS3 from VGGSound. We also design a Sound Source Matching Score to clearly measure localized audio relevance. With the effectiveness of explicit sound source modeling, SS2A achieves state-of-the-art performance in extensive image-to-audio tasks. We also qualitatively demonstrate SS2A's ability to achieve intuitive synthesis control by compositing vision, text, and audio conditions. Furthermore, we show that our sound source modeling can achieve competitive video-to-audio performance with a straightforward temporal aggregation mechanism. 
\end{abstract}

\begin{links}
    \link{Demo Website}{https://SSV2A.github.io/SSV2A-demo/}
\end{links}

\section{Introduction}  
As multimedia consumption surges, generating sound for a silent scene attracts high demands in various industries \cite{applications}. The synthesized audio can complement a virtual reality scene \cite{VR}, create Foley for films and games \cite{foley}, and sonify visual contents for people with visual impairment \cite{zhou2018wild}. By learning from text-audio or visual-audio pairs, recent methods can generate highly relevant audio clips given conditions as texts, images or videos. However, most existing methods \cite{AudioGenSurvey} only model the mapping between global visual scene and sound while overlooking local details.

\begin{figure}
    \centering
    \includegraphics[width=\linewidth]{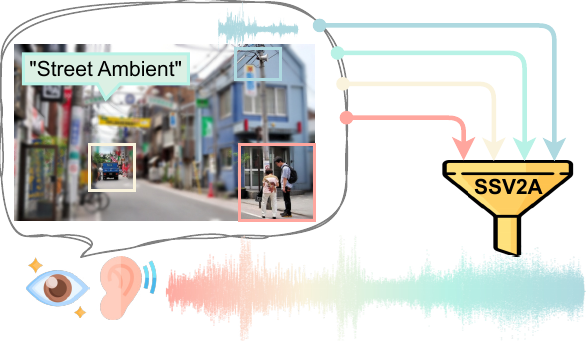}
    \caption{Our SS2A perceives multimodal sound sources in a scene for V2A immersiveness and expressiveness.}
    \label{fig:teaser}
\end{figure}

In reality, sound is produced and recognized from sounding objects, \textit{i.e.}, \textbf{sound sources}, locally present in a soundscape \cite{SoundRecog}. For instance, in a street the sound comes from individual vehicles and passengers as illustrated in \cref{fig:teaser}. Humans also perceive audio immersiveness and expressiveness from sound source interactions \cite{VisualAud}. In practice, audio engineers leverage sound sources to intuitively synthesize sounds \cite{SoundSynthesis}.

\begin{figure*}
    \centering
        \includegraphics[width=\linewidth]{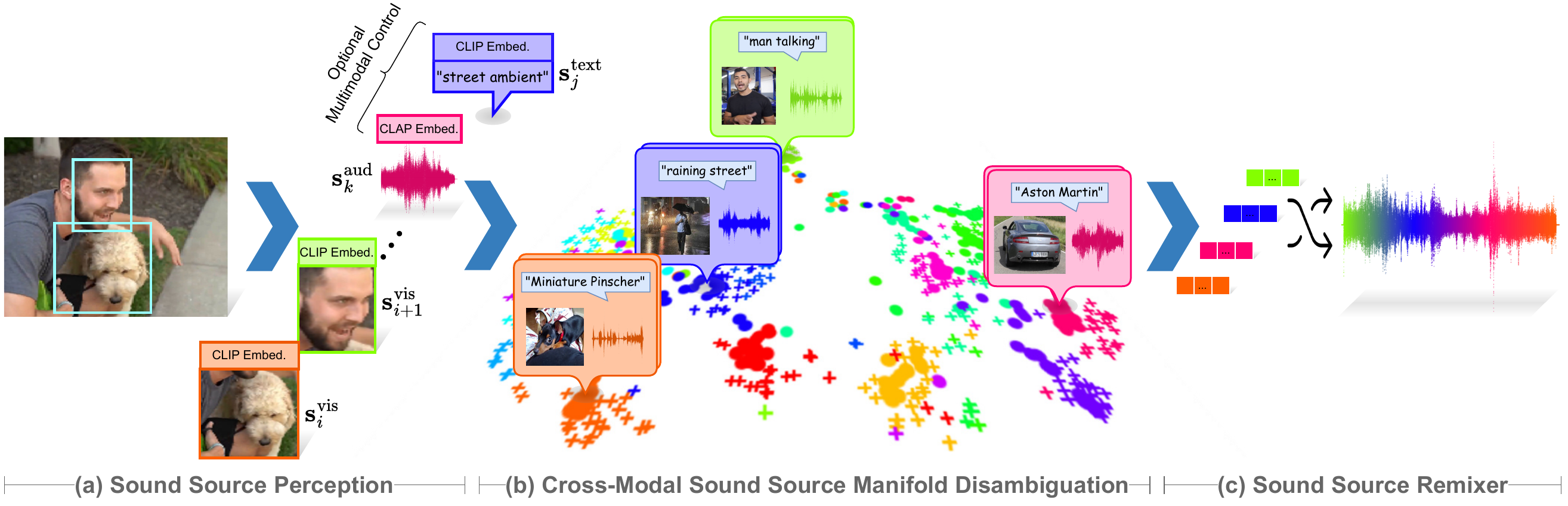}
        \caption{\textbf{Pipeline of SS2A.} We perceive sound sources prompted by vision, text, or audio and disambiguate them in the semantically learned CMSS Manifold, which are then mixed to generate an audio clip with immersiveness and expressiveness.}
        \label{fig:pipeline}
\end{figure*}

Can an audio synthesizer utilize \textbf{sound source-aware} conditions to obtain better generation quality and control? To answer this question, we present a \textbf{S}ound \textbf{S}ource-Aware \textbf{A}udio (\textbf{SS2A}) generator. As image offers sound sources with straightforward curation and composition, we choose it as the primary modality to condition SS2A. We model our system in semantic spaces for learning efficiency and include multimodal conditions from text and audio to boost sound source control. As depicted in \cref{fig:teaser}, the perception can also come from audio sound source as a loudspeaker and text source as ``street ambient". We present SS2A's pipeline in \cref{fig:pipeline}. SS2A first \textbf{perceives} multimodal sound source conditions as CLIP \cite{CLIP} or CLAP \cite{CLAP} semantic embeddings with visual detection and cross-modal translation. We then project them to a Cross-Modal Sound Source (CMSS) Manifold to \textbf{disambiguate} each source. By disambiguation, we require the CMSS manifold to (1) contrast the source semantics and (2) respect the audio characteristics of each sound source. After querying CMSS embeddings of individual sound sources, SS2A learns an attention-based Sound Source Remixer to \textbf{mix} them into a CLAP audio embedding with rich sound source information. This representation is passed to a pretrained audio generator, AudioLDM \cite{AudioLDM}, to synthesize the output audio waveform.

As the CMSS manifold contrastively learns from single-sound-source image-audio pairs to disambiguate sound source semantics, we filter the VGGSound \cite{VGGS} data with visual detection to form a novel dataset, VGGSound Single Source (VGGS3), that contains 106K high-quality single-sound-source image-audio pairs. We also apply a novel Cross-Modal Contrastive Mask Regularization (CCMR) during manifold learning to retain rich CLIP-CLAP semantics by reducing CMSS contrastive influence on similar visual-audio sources with CLIP and CLAP priors. To effectively evaluate generation relevance, we introduce a Sound Source Matching Score (SSMS) to compute the F1 score of overlapping sound source labels on ground-truth and generated samples with an audio classifier.

Both objective and subjective results show that SS2A achieves state-of-the-art performance in image-to-audio synthesis, indicating the benefits of sound source modeling. We demonstrate SS2A's intuitive generation control by flexibly compositing multimodal sound source prompts from vision, text, and audio to synthesize immersive qualitative samples. We further showcase that our sound source modeling can be straightforwardly extended to competitive video-to-audio synthesis with a temporal aggregation mechanism.

In summary, our contributions are as follows:

\begin{itemize}
    \item We present a novel framework, SS2A, addressing audio synthesis at the sound-source level. Extensive experiments show that our multimodal sound source modeling leads to state-of-the-art results in image-to-audio generation and competitive video-to-audio performance.
    \item We explore how sound-source disambiguation can enhance SS2A synthesis with the CMSS manifold, along with a novel CCMR mechanism to guide cross-modal contrastive learning with foundation model priors.
    \item During manifold training, we curate a high-quality single-sound-source image-audio dataset, VGGS3.
    \item In evaluating relevance between generated and ground-truth audio signals, we introduce a novel SSMS metric to explicitly match their localized sound sources, proposing a new objective for fine-grained audio generation.
    \item We showcase multimodal sound source composition, a fresh audio synthesis paradigm that offers intuitive generation control over a wide range of usage scenarios.
\end{itemize}

\begin{figure*}[!ht]
    \centering
    \includegraphics[width=\linewidth]{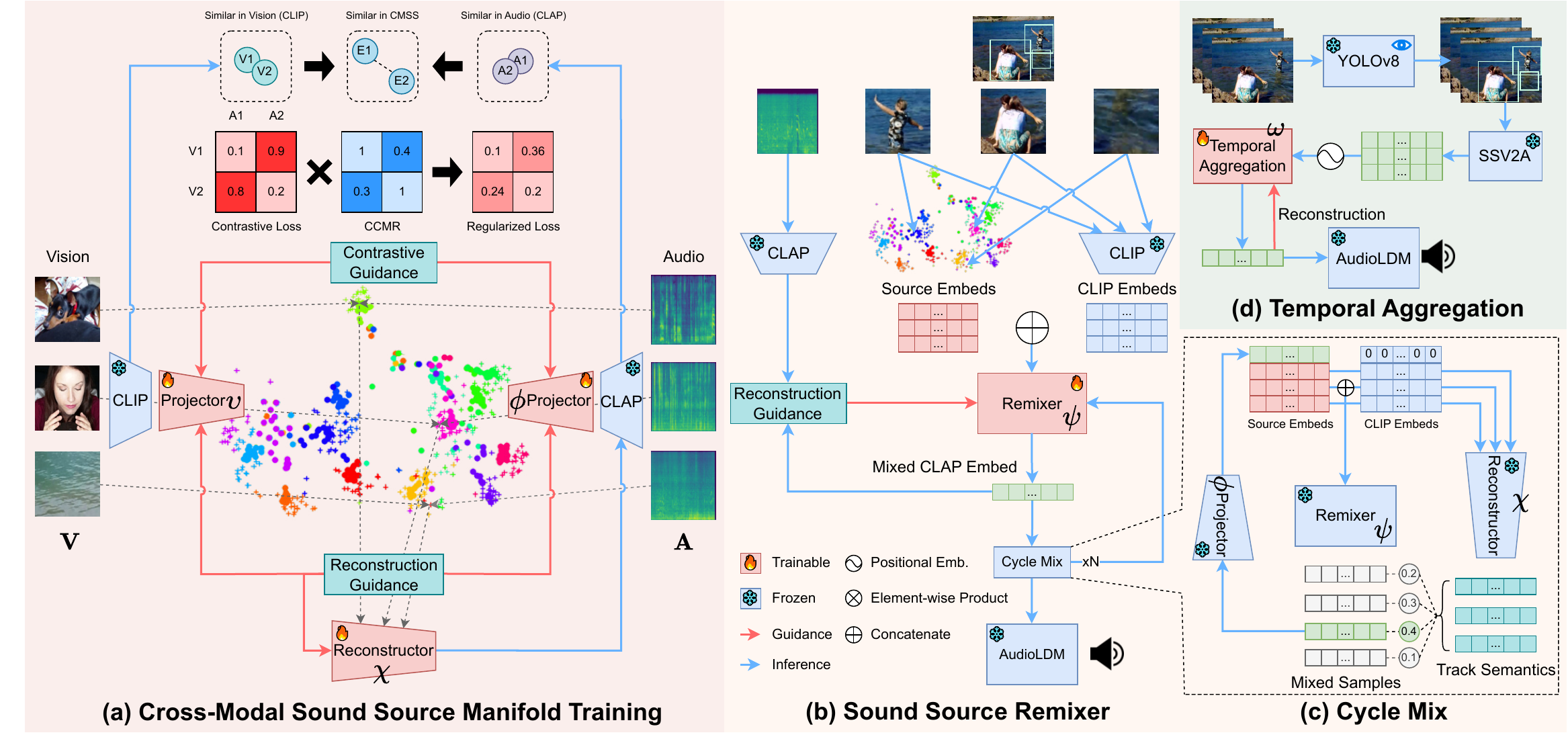}
    \caption{\textbf{Detailed Schematics of SS2A Modules.} \textbf{(a)} We learn two projectors to map the CLIP-CLAP embeddings of single-source visual-audio pairs to a joint semantic space with contrastive guidance, forming our CMSS manifold. An auxiliary CLAP reconstruction encodes audio semantics into this manifold. \textbf{(b)} The Sound Source Remixer attends to the CMSS embeddings concatenated with their CLIP semantics, generating a single CLAP audio representation which is passed to AudioLDM. \textbf{(c)} We reuse the CMSS reconstructor to generate source-wise ``track semantics" in CLAP space and refine the Remixer samples iteratively. \textbf{(d)} We train an additional Temporal Aggregation (TA) module to attend to positionally embedded SS2A generations across video frames and enhance visual-audio synchronization.}
    \label{fig:schematics}
\end{figure*}

\section{Related Works}

\subsection{Vision-to-Audio Generation}
Early V2A methods \cite{owens2016vis, chen2017deep, zhou2018wild, hao2018cmcgan, chen2018classvis, chen2020gan} train a source-specific V2A model on each audio class and cannot generalize to open-domain V2A synthesis. As a precursor, recent SpecVQGAN \cite{SpecVQGAN} learns a discrete neural codec \cite{VQVAE, VQGAN} of source-agnostic audio features and autoregressively generates audio codes with a Transformer \cite{transformer}. Following SpecVQGAN, Im2Wav \cite{Im2Wav} further details its audio codec into low-level and high-level features. MaskVAT \cite{MaskVAT} leverages a pretrained codec DAC \cite{DAC} and predicts audio tokens with a Masked Generative Transformer \cite{MaskGIT}. Another line of methods employ Diffusion \cite{diffusion} models. CLIPSonic-IQ \cite{CLIPSonic} queries CLIP \cite{CLIP} to condition its Diffusion process. Diff-Foley \cite{Diff-Foley} contrastively learns a temporally-aligned visual-audio prior to guide video-audio synchronization. Draw-an-Audio \cite{DrawAudio} leverages loudness signal, text caption, and masked video conditions simultaneously. More recently, some methods bridge visual conditions to the prior of a pretrained audio generator for efficient V2A learning. V2A-Mapper \cite{V2A-Mapper} maps CLIP embeddings to CLAP \cite{CLAP} space, from which a pretrained AudioLDM \cite{AudioLDM} model synthesizes the audio signal. V2A-SceneDetector \cite{Scene-V2A-Mapper} extends V2A-Mapper to multi-scene video with a detection module. Seeing and Hearing \cite{SandH} aligns ImageBind \cite{imagebind} visual embeddings to AudioLDM. FoleyCrafter \cite{FoleyCrafter} devises a timestamp predictor to enhance synchronization during bridging. Very recently, FRIEREN \cite{frieren} and MMAudio \cite{MMAudio} explore V2A generation with Rectified Flow Matching \cite{rfm}. MultiFoley \cite{MultiFoley} employs a diffusion transformer to jointly map multimodal conditions to audio. Most existing methods condition on global visual scenes for V2A synthesis. Some recent works \cite{CycV2A} \cite{ObjectAware} leverage pixel-level conditions for V2A synthesis, partially describing visual sounding objects. In reality, human perceive object-level sound sources across modalities and time \cite{SoundRecog}. Such a sound source-aware V2A generator remains uninvestigated.

\subsection{Contrastive Cross-Modal Alignment}
Contrastive representation learning \cite{contrast} has significantly advanced cross-modal representation alignment. CLIP \cite{CLIP} aligns text and image modalities by learning from abundant text-image pairs. Many aforementioned V2A methods \cite{Im2Wav, MaskVAT, CLIPSonic, FoleyCrafter, V2A-Mapper} benefit from its semantically rich visual representations. Similarly, CLAP \cite{CLAP} learns from text-audio pairs and is used extensively in V2A generation \cite{Diff-Foley, SandH, V2A-Mapper, DrawAudio, AudioLDM, AudioLDM2}. Aside from modality alignment, Diff-Foley \cite{Diff-Foley} shows that it is possible to respect temporal alignment in the contrastive visual-audio representation to benefit video-audio synchronization. However, the entanglement of temporal features in this representation limits Diff-Foley in generalizing to image-to-audio synthesis. In this work, we focus on taming a contrastive representation for sound source disambiguation and leave the temporal alignment to a downstream temporal aggregation module.

\section{Method} \label{sec:method}

Approximating an audio distribution $Q(\mathbf{A}\vert\mathbf{a})$, the audio generator AudioLDM \cite{AudioLDM} generates audio signals $\mathbf{A}$ from CLAP \cite{CLAP} audio semantics $\mathbf{a}$. For learning efficiency, we employ a pretrained $Q$ and synthesize $\mathbf{a}$ instead of $\mathbf{A}$. Conditioned on multimodal sound sources, our objective is to learn a conditional distribution:

\begin{equation}\label{eq:p}
    P \left( \mathbf{a} \ \vert \ \left\{\mathbf{s}^\text{vis}_i\right\},\left\{\mathbf{s}^\text{text}_j\right\},\left\{\mathbf{s}^\text{aud}_k\right\} \right) ,
\end{equation}

\noindent where $\left\{\mathbf{s}^\text{vis}_i\right\}$, $\left\{\mathbf{s}^\text{text}_j\right\}$, and $\left\{\mathbf{s}^\text{aud}_k\right\}$ denote respectively the semantic embedding sets of $I$ visual sound sources, $J$ text sources and $K$ audio sources encoded with CLIP \cite{CLIP} or CLAP. We term the acquisition of these semantics as Sound Source Perception in \cref{fig:pipeline} (a).

The most straightforward way to approximate \cref{eq:p} is to train a standalone model that maps the perceived CLIP-CLAP semantics directly to $\mathbf{a}$. However, two CLIP features \textbf{ambiguate} this direct learning: (1) the CLIP image space models global visual context rather than contrasting individual objects, and (2) CLIP learns only from text-image data, which lacks awareness of the sources’ audio traits.  As an efficient solution, we learn a Cross-Modal Sound Source (CMSS) manifold as illustrated in \cref{fig:pipeline} (b) to project the CLIP-CLAP embeddings to a joint semantic space where the local sound sources are \textbf{disambiguated}.

Finally, we attentively mix the CMSS embeddings together in \cref{fig:pipeline} (c) to generate $\mathbf{a}$. This stage involves an attention-based Sound Source Remixer module.

\subsection{Sound Source Perception} \label{subsec:perceive}
Recall Eq. \eqref{eq:p}. To extract $\left\{\mathbf{s}^\text{vis}_i\right\}$ from a global visual cue when no manual sound-source annotation is available, we pass each image through a visual detector and crop out the detected regions with predicted bounding boxes. These image regions are then embedded by CLIP. To obtain $\left\{\mathbf{s}^\text{text}_j\right\}$, we translate the CLIP text embeddings of text prompts to CLIP image space with a pretrained DALL$\cdot$E-2 Prior \cite{DALLE2} model to mitigate the visual-text domain gap \cite{gap} and ease downstream disambiguation. For $\left\{\mathbf{s}^\text{aud}_k\right\}$, we pass the audio prompts through CLAP to get embeddings.

\subsection{Cross-Modal Sound Source Manifold} \label{subsec:manifold}
We contrastively learn the CMSS manifold from single-sound-source visual-audio pairs to project the perceived sound source semantics to a joint semantic space for disambiguation, as shown in \cref{fig:schematics} (a). The CMSS manifold naturally accommodates the multimodality of our perceptions due to the bridging of CLIP and CLAP.

\paragraph{Manifold Learning.}
We formulate two CMSS manifold projections $\upsilon\left(\cdot\right)$ and $\phi\left(\cdot\right)$ as:

\begin{equation}
    \mathbf{e}_\text{CLIP}=\upsilon \left( \mathbf{v} \right), \ \mathbf{e}_\text{CLAP}=\phi \left(\mathbf{a} \right),
\end{equation}

\noindent given a single-source visual-audio pair as $\left(\mathbf{V}, \mathbf{A}\right)$ and its CLIP-CLAP embeddings as $\left(\mathbf{v}, \mathbf{a}\right)$. $\mathbf{e}$ denotes the CMSS embedding. The projectors optimize a contrastive loss to attract visual-audio embeddings from the same sound-source pair and repel those from different sources. Following the symmetric contrastive guidance of CLAP \cite{CLAP}, this objective can be formulated for a batch of $N$ pairs as:

\begin{equation}\label{eq:Lc}
    \mathcal{L}_c=\frac{\ell_\text{CLIP}\left(\mathbf{C}\right)+\ell_\text{CLAP}\left(\mathbf{C}\right)}{2},
\end{equation}

\noindent where $\ell_\text{CLIP}\left(\mathbf{C}\right)=\frac{1}{N}\sum_{i=0}^{N}\log diag \left(softmax \left(\mathbf{C}\right)\right)$ penalizes off-diagonal similarities in similarity entries $\mathbf{C}_{ij}=\tau \ast \left[\mathbf{e}^i_\text{CLIP} \cdot (\mathbf{e}_\text{CLAP}^{j})^\top\right]$. $\ell_\text{CLAP}$ follows $\ell_\text{CLIP}$ but swaps $\mathbf{e}_\text{CLIP}$ and $\mathbf{e}_\text{CLAP}$ in $\mathbf{C}_{ij}$. $\tau$ is a learned temperature parameter.

We define an auxiliary reconstruction $\chi\left(\cdot\right)$ to map the CMSS embeddings back to CLAP space, assisting their alignment with audio semantics. The reconstruction objective is designated for each visual-audio pair as:

\begin{equation}
    \mathcal{L}_r = \frac{\lVert 1 - sim(\mathbf{a},\  \chi(\mathbf{e}_\text{CLAP})) \rVert + \lVert 1 - sim(\mathbf{a},\  \chi(\mathbf{e}_\text{CLIP})) \rVert}{2},
\end{equation}

\noindent where $sim\left(\cdot,\cdot\right)$ computes the cosine similarity.

We model $\upsilon\left(\cdot\right)$, $\phi\left(\cdot\right)$, and $\chi\left(\cdot\right)$ variationally with the reparameterization trick and add a Kullback-Leibler (K-L) divergence regularization term to each against the standard normal distribution as $\mathcal{L}_{kl}$. The final objective is then:

\begin{equation}
    \mathcal{L}_\text{fold}=\mathcal{L}_c + \mathcal{L}_r + \lambda_1\mathcal{L}_{kl},
\end{equation}

\noindent where $\lambda_1$ is a weight hyperparameter and $\mathcal{L}_{kl}$ is the summed K-L loss. During training, we model all three modules $\upsilon\left(\cdot\right)$, $\phi\left(\cdot\right)$, and $\chi\left(\cdot\right)$ with residually connected MLPs and alternatively optimize the projectors and generator with $\mathcal{L}_\text{fold}$.

\paragraph{Cross-Modal Contrastive Mask Regularization.}
To avoid the loss of rich semantics from CLIP and CLAP due to small training data, we employ a Cross-Modal Contrastive Mask Regularization (CCMR) mechanism to weaken the contrastive guidance $\mathcal{L}_c$ defined in \cref{eq:Lc} for similar cross-pair audio-visual samples. For each batch, we compute a CLIP-CLIP similarity matrix $\mathbf{M}_\text{CLIP}$ and a CLAP-CLAP similarity matrix $\mathbf{M}_\text{CLAP}$ per entry as:

\begin{equation}
    \mathbf{M}_\text{CLIP}^{ij}=sim\left(\mathbf{v}_i, \mathbf{v}_j\right), \  \mathbf{M}_\text{CLAP}^{ij}=sim\left(\mathbf{a}_i, \mathbf{a}_j\right).
\end{equation}

\noindent The CCMR mask $\mathbf{M}$ is then computed per entry as:

\begin{equation} \label{eq:ccmr}
    \mathbf{M}_{ij}=e^{-\alpha \ \ast \ \left( clamp\left(\mathbf{M}_\text{CLIP}^{ij} \ \ast \ \mathbf{M}_\text{CLAP}^{ij}\right)\right)^\alpha},
\end{equation}

\noindent where $clamp\left(\cdot\right)$ restricts the mask entry to be within $\left[0,1\right]$. This is a stretched exponential decay that grows smaller when both $\mathbf{M}^{ij}_\text{CLIP}$ and $\mathbf{M}^{ij}_\text{CLAP}$ increase. The hyperparameter $\alpha$ controls the decay curvature and steepness. We apply $\mathbf{M}$ to the original contrastive similarity matrix $\mathbf{C}$ with an element-wise multiplication as $\mathbf{C}^\ast_{ij}=\mathbf{C}_{ij}\ast\mathbf{M}_{ij}$.

\paragraph{Data Curation and Training.} \label{par:data_curation}
We filter visual-audio pairs from VGGSound \cite{VGGS} training set with a visual detection pipeline and obtain 106K single-sound-source visual-audio pairs as a novel dataset VGGSound Single Source (VGGS3). We term the VGGS3 pairs \textit{curated pairs}. Additionally, we translate the single-source text-audio pairs from LAION-630K \cite{LAION630K} to visual-audio pairs with a pretrained DALL$\cdot$E-2 Prior \cite{DALLE2} model. We term these pairs \textit{translated pairs}. A Mean-Teacher \cite{MeanTeacher} paradigm trains the CMSS modules with these pairs. Please refer to Supplementary Section 2 for our data curation and training details.

\begin{table*}
\centering
\fontsize{9pt}{9pt}\selectfont
\begin{tabular}{l | l l l l | l l | l l}
    \toprule
      \multirow{2}{*}[-2pt]{Method} & \multicolumn{4}{c|}{VGGSound} & \multicolumn{2}{c|}{Subjective Relevance} & \multicolumn{2}{c}{Subjective Fidelity}\\ \cmidrule(l{5pt}r{5pt}){2-5} \cmidrule(l{5pt}r{5pt}){6-7} \cmidrule(l{5pt}r{5pt}){8-9}
      & V-FAD$\downarrow$ & C-FAD$\downarrow$ &  CS$\uparrow$ & SSMS$\uparrow$ & MOS$\uparrow$ & Std. & MOS$\uparrow$ & Std.\\
      \midrule
      S\&H & 10.929 & 79.221 & 5.941 & 1.886 & - & - & - & -\\
      S\&H-Text & 5.087 & 31.461 & 8.690 & 2.646 & 2.358 & 0.845 & 2.460 & 0.748\\
      Im2Wav & 5.855 & 24.303 & 10.965 & 3.105 & 2.595 & 0.736 & 2.273 & 0.757\\
      V2A-Mapper & \textbf{0.946} & \textbf{5.516} & \underline{11.521} & \underline{3.164} & \underline{3.063} & 1.024 & \underline{2.693} & 1.210\\ \cmidrule(l{5pt}r{5pt}){1-1} \cmidrule(l{5pt}r{5pt}){2-5} \cmidrule(l{5pt}r{5pt}){6-7} \cmidrule(l{5pt}r{5pt}){8-9}
      SS2A(Ours) & \underline{1.150} & \underline{6.716} & \textbf{12.947} & \textbf{3.793} & \textbf{4.080} & 0.527 & \textbf{4.098} & 0.459\\
    \bottomrule
\end{tabular}
\captionof{table}{\textbf{General image to audio tests.} The VGGSound and subjective tests, without source annotation, generalize single-source and multi-source synthesis scenarios. The first and second places are \textbf{bolded} and \underline{underlined}, respectively.}
\label{table:objective-subjective}
\end{table*}

\subsection{Sound Source Remixer} \label{subsec:remixer}
We employ a Sound Source Remixer function $\psi\left(\cdot\right)$ to mix the embeddings $\left\{\mathbf{e}_m\right\}$ queried from the CMSS manifold in \cref{fig:schematics} (b), generating a CLAP audio representation with rich sound source semantics as $\mathbf{a}_\text{mix}$. To leverage all the semantical features helpful for this task, we concatenate each $\mathbf{e}$ with its CLIP embedding $\mathbf{v}$. Specifically, given a set of $M$ sound sources, we formulate $f_\text{mix}$ as:

\begin{equation} \label{eq:mix}
    \psi\left(\mathbf{x}_1, \mathbf{x}_2, \cdots, \mathbf{x}_M\right)=\mathbf{a}_\text{mix},
\end{equation}

\noindent where $\mathbf{x}_i=concat(\mathbf{e}_m, \mathbf{v}_m)$ is the concatenated token for the $m$-th source. We model $\psi\left(\cdot\right)$ variationally to make it generative. The optimization objective is designed as:

\begin{equation}
    \mathcal{L}_\text{mix}=\lVert1-sim(\mathbf{a}, \mathbf{a}_\text{mix})\rVert+\lambda_2 \mathcal{L}_{kl},
\end{equation}

\noindent where $\mathcal{L}_{kl}$ is the KL divergence from standard normal distribution, and $\lambda_2$ is a weight hyperparameter.

We model $\psi\left(\cdot\right)$ with a stack of self-attention layers and learn it from visual-audio pairs in VGGSound. The visual sources are perceived from each video's central frame following our aforementioned perception method. Each token sequence $\left\{\mathbf{x}_m\right\}$ is padded to a fixed length of $M=64$. To enhance generation diversity, a Classifier-free Guidance \cite{CFG} is applied during training by randomly zeroing out tokens. We replace the classic attention with Efficient Attention \cite{EfficientAtt} and detail this architecture in Supplementary Section 2.3. During inference, we set $\mathbf{v}=\mathbf{0}$ for sound source conditions from audio modality.

\paragraph{Cycle Mix.}
We can also obtain a CLAP embedding $\mathbf{a}_\text{src}=\chi\left(\mathbf{e}\right)$ for each sound source through the CMSS manifold's reconstructor. $\mathbf{a}_\text{src}$ can be regarded as a set of source-wise audio semantics generated by our method. As one of our objectives for $\mathbf{a}_\text{mix}$ is to have high relevance to each sound source, $\left\{\mathbf{a}^m_\text{src}\right\}$ are recycled to iteratively guide the generation of $\mathbf{a}_\text{mix}$. This mechanism, termed Cycle Mix, is illustrated in \cref{fig:schematics} (c) and Algorithm 1 in Supplementary Section 2.3.

\paragraph{Temporal Aggregation.}
So far, the Sound Source Remixer learns an image-to-audio task. Following V2A-Mapper \cite{V2A-Mapper}, we adapt it to the video-to-audio task with a downstream Temporal Aggregation (TA) function $\omega\left(\cdot\right)$ depicted in \cref{fig:schematics} (d). Instead of averaging the frame-wise semantics, we learn a nonlinear $\omega\left(\cdot\right)$. We evenly extract 64 frames along time from one video and generate a CLAP embedding for each of them. Each embedding is then positionally embedded with its timestamp. $\omega\left(\cdot\right)$ learns to fuse these embeddings into a temporally-aligned CLAP audio representation $\mathbf{a}$ with the following loss:

\begin{equation}
    \mathcal{L}_\text{ta}=\lVert 1 - sim\left(\mathbf{a}, \omega\left(pos\left(\mathbf{a}^1_\text{gen}, \cdots, \mathbf{a}^{64}_\text{gen}\right), t \right) \right) \rVert,
\end{equation}

\noindent where $\mathbf{a}_\text{gen}$ denotes the SS2A generated CLAP embeddings and $pos\left(\cdot, t\right)$ is the positional embedding function. The architecture of TA is a stack of self-attention layers.

\begin{table*}
\centering
\fontsize{9pt}{9pt}\selectfont
\begin{tabular}{l | l | l l l l | l l l l | l}
    \toprule
      & \multirow{2}{*}[-2pt]{Method} & \multicolumn{4}{c|}{VGG-SS} & \multicolumn{4}{c|}{MUSIC} & ImageHear \\ \cmidrule(l{5pt}r{5pt}){3-6} \cmidrule(l{5pt}r{5pt}){7-10} \cmidrule(l{5pt}r{5pt}){11-11}
      & & V-FAD$\downarrow$ & C-FAD$\downarrow$ &  CS$\uparrow$ & SSMS$\uparrow$ & V-FAD$\downarrow$ & C-FAD$\downarrow$ &  CS$\uparrow$ & SSMS$\uparrow$ & CS$\uparrow$\\
     \midrule
     \multirow{8}{*}{\rotatebox{90}{Single-Source}} & GroundTruth & 0 & 0.171 & 13.199 & 10 & 0 & 0 & 13.906 & 10 & -\\
     & Oracle & 1.400 & 9.983 & 12.071 & 5.752 & 6.430 & 25.422 & 12.861 & 7.777 & -\\ \cmidrule(l{5pt}r{5pt}){2-2} \cmidrule(l{5pt}r{5pt}){3-6} \cmidrule(l{5pt}r{5pt}){7-10}
     & S\&H & 16.015 & 90.656 & 5.901 & 1.903 & 49.045 & 156.898 & 4.126 & 1.421 & 3.417\\
     & S\&H-Text & 7.118 & 37.899 & 9.761 & 3.685 & 25.081 & 77.218 & 10.259 & 5.635 & 7.401\\
     & Im2Wav & 7.573 & 29.213 & 11.011 & 4.451 & 26.344 & 57.596 & 8.374 & 6.214 & 10.758\\
     & RAM+ALDM & 6.532 & 30.461 & 9.199 & 2.714 & 23.681 & 63.810 & 7.795 & 3.421 & 8.765\\
     & V2A-Mapper & \textbf{1.666} & \textbf{13.583} & \underline{11.842} & \underline{4.488} & \textbf{7.245} & \underline{27.657} & \underline{12.901} & \underline{6.288} & \underline{12.689}\\ \cmidrule(l{5pt}r{5pt}){2-2} \cmidrule(l{5pt}r{5pt}){3-6} \cmidrule(l{5pt}r{5pt}){7-10} \cmidrule(l{5pt}r{5pt}){11-11}
     & SS2A (Ours) & \underline{2.815} & \underline{15.150} & \textbf{12.215} & \textbf{4.936} & \underline{8.075} & \textbf{25.390} & \textbf{13.859} & \textbf{7.330} & \textbf{13.930}\\ \cmidrule(l{5pt}r{5pt}){1-2} \cmidrule(l{5pt}r{5pt}){3-6} \cmidrule(l{5pt}r{5pt}){7-10} \cmidrule(l{5pt}r{5pt}){11-11}
     \multirow{8}{*}{\rotatebox{90}{Multi-Source}} & GroundTruth & 0  & 0.793 & 12.344 & 10 & 0 & 0 & 13.009 & 10 & -\\
     & Oracle & 4.356 & 31.569 & 11.840 & 6.447 & 1.492 & 34.295 & 11.658 & 6.300 & -\\ \cmidrule(l{5pt}r{5pt}){2-2} \cmidrule(l{5pt}r{5pt}){3-6} \cmidrule(l{5pt}r{5pt}){7-10}
     & S\&H & 21.447 & 121.371 & 6.594 & 2.568 & 27.661 & 175.708 & 3.979 & 0.986 & -\\
     & S\&H-Text & 12.678 & 81.944 & 9.573 & 4.026 & 9.887 & 105.529 & 9.149 & 5.223 & -\\
     & Im2Wav & 12.915 & 64.648 & 11.309 & \underline{5.132} & 12.055 & 81.321 & 6.426 & \underline{5.357} & -\\
     & RAM+ALDM & 14.820 & 76.406 & 9.009 & 3.026 & 12.985 & 92.316 & 8.892 & 4.261 & -\\
     & V2A-Mapper & \underline{10.228} & \underline{59.660} & \underline{11.331} & 4.684 & \underline{4.490} & \underline{48.665} & \underline{11.126} & 4.907 & -\\ \cmidrule(l{5pt}r{5pt}){2-2} \cmidrule(l{5pt}r{5pt}){3-6} \cmidrule(l{5pt}r{5pt}){7-10}
     & SS2A (Ours) & \textbf{6.810} & \textbf{46.933} & \textbf{11.744} & \textbf{5.973} & \textbf{3.387} & \textbf{31.115} & \textbf{12.951} & \textbf{6.000} & -\\
     \bottomrule
\end{tabular}
\caption{\textbf{Source-annotated image to audio tests.} These datasets have source annotations to differentiate single-source and multi-source generation scenarios. Only CS is available for ImageHear as it lacks ground-truth pairing audio with each image.}
\label{table:source-objective}
\end{table*}

\section{Experiments and Results}

\subsection{Experimental Setup}
Please see Supplementary Section 2 for SS2A's implementation details along with the architecture designs.

\smallskip
\noindent
\textbf{Datasets.}
We train our teacher CMMS manifold modules on the VGG Sound Source (VGG-SS) \cite{VGG-SS} dataset. The student modules learn from (1) VGG-SS and (2) curated and translated visual-audio pairs. Since VGG-SS does not have an official train-test split, we randomly sample 4.5K pairs from it for training and form a test set with the remaining 500 pairs. We train the Sound Source Remixer modules following the provided train-test split on VGGSound \cite{VGGS}, which contains 19K pairs across 310 audio categories. For image to audio tasks, we test on the VGGSound test set excluding VGG-SS entries, generating 10288 samples. This test does not differentiate single-source and multi-source generation scenarios as VGGSound has no source annotations. For source-annotated tests that clearly split these scenarios, we focus on VGG-SS which contains 38 multi-source pairs (2\~10 sources each) and 455 single-source pairs. We also test on two out-of-distribution sets MUSIC \cite{MUSIC} and ImageHear \cite{Im2Wav} to show SS2A's generalization capability. MUSIC contains 140 pairs with duet musical instrument performance, and 1034 pairs with solo instrument. ImageHear has 101 single-source images from 30 visual classes. We generate 10-second audio samples for all tests.

\smallskip
\noindent
\textbf{Objective Metrics.}
We measure generation quality objectively from two perspectives: fidelity and relevance. For generation fidelity, we adopt the Fr\'echet Audio Distance (FAD) \cite{FAD} with an open-source implementation \cite{fad-git} to obtain two metrics, V-FAD and C-FAD, respectively from VGGish \cite{FAD} and CLAP \cite{CLAP} models. FAD measures the closeness of ground-truth and generated audio feature distributions. A low FAD score reflects high generation fidelity. For generation relevance, we adopt the CLIP-Score (CS) which maps an audio's CLAP embedding to the CLIP image space with a Wav2CLIP \cite{Wav2CLIP} model to compare its similarity with the paired image. For multi-source image-audio pairs in VGG-SS, we average CS between each sound source image and the paired audio. We compute CS on global images in other tests. A high CS represents high generation relevance.

\smallskip
\noindent
\textbf{Matching Score.}
We observe that the CS relevance comparison, by mapping audio features to image domain, causes loss of audio information. As a result, our method often outperforms Oracle AudioLDM generations in CS scoring from \cref{table:source-objective}. We propose a novel metric, Sound Source Matching Score (SSMS), that adopts an audio classifier BEATs \cite{beats} to respectively predict $N$ localized sound source labels for ground-truth and generated audios. We regard intersected labels from both sets as true positives, the difference of ground-truth against generation as false negatives, and the reverse difference as false positives. SSMS is computed as the F1 score of these statistics. We set $N=10$ throughout experiments and show that SSMS distinguishes generation relevance more clearly than CS.

\smallskip
\noindent
\textbf{Subjective Metrics.}
Following recent works \cite{V2A-Mapper, FoleyCrafter}, we conduct a subjective listening test with 20 human evaluators. We randomly sample 40 central video frames from AudioSet Strong \cite{audiosetstrong} and AVSBench \cite{AVSBench}, generating 10-second audio clips with each image-to-audio method. The participants are asked to rate 20 of them for fidelity without visual cues. They then rate 20 samples for relevance given the visual conditions. We collect the ratings on a 5-point scale and compute the Mean Opinion Score (MOS) \cite{MOS} to measure generation fidelity and relevance. Please see Supplementary Section 6 for the evaluation setup.

\subsection{Baseline Evaluations}
We compare our generator with three image-to-audio methods: V2A-Mapper \cite{V2A-Mapper}, Seeing and Hearing (S\&H) \cite{SandH}, and Im2Wav \cite{Im2Wav}. Additionally, we employ RAM \cite{RAM} to generate image tags and pass them to GPT-4 \cite{GPT4} for text captions, which are fed to AudioLDM to generate audio. We call this cascaded baseline RAM+ALDM. We qualitatively demonstrate how cascaded methods are inferior to SS2A in Supplementary Section 7.

For video-to-audio tasks, we compare with Diff-Foley \cite{Diff-Foley}, Frieren \cite{frieren}, MultiFoley \cite{MultiFoley}, and MMAudio \cite{MMAudio}. Some baselines require different visual conditions. For fairness, we modify some methods following Supplementary Section 1 but still keep their original results.

\smallskip
\noindent
\textbf{Objective Results.} As illustrated in \cref{table:objective-subjective} and \cref{table:source-objective}, our method achieves superior performance in most objective metrics for both in-distribution and out-of-distribution tests. For single-source generation, we outperform baselines in generation relevance and stay in top 2 for generation fidelity. For multi-source generation, SS2A is superior in all metrics. Surprisingly, SS2A achieves a higher CS in generation relevance than the Oracle baseline, which is assumed to have optimal performance for V2A methods involving AudioLDM. This effect is no longer observed in SSMS, demonstrating our new metric's superiority in comparing audio generation relevance. Even S\&H-Text has seen generated text captions, SS2A still surpasses it in both fidelity and relevance.

SS2A performs competitively in video-to-audio tasks with the TA extension as shown in Supplementary Section 5.1, showing that our sound source modeling can also benefit video-to-audio synthesis with a straightforward temporal feature integration.

\smallskip
\noindent
\textbf{Subjective Results.} In \cref{table:objective-subjective}, our method outperforms baselines significantly in human-evaluated generation fidelity and relevance. We choose to test S\&H-Text instead of S\&H to obtain the best generation performance Seeing and Hearing can achieve, even though it sees extra text captions.

\begin{table*}
    \fontsize{9pt}{9pt}\selectfont
    \centering
    \begin{tabular}{c c | l l l l | l l l l}
    \toprule
    \multirow{2}{*}[-2pt]{CMSS} & \multirow{2}{*}[-2pt]{CLIP} & \multicolumn{4}{c|}{Single-Source Generation} & \multicolumn{4}{c}{Multi-Source Generation} \\ \cmidrule(l{5pt}r{5pt}){3-6} \cmidrule(l{5pt}r{5pt}){7-10}
    & & V-FAD$\downarrow$ & C-FAD$\downarrow$ & CS$\uparrow$ & SSMS$\uparrow$ & V-FAD$\downarrow$ & C-FAD$\downarrow$ & CS$\uparrow$ & SSMS$\uparrow$\\
    \midrule
    \xmark & \cmark & 39.622 & 122.127 & 3.987 & 1.385 & 34.378 & 119.692 & 5.574 & 1.579\\
    \cmark & \xmark & 17.949 & 96.045 & 6.049 & 1.213 & 7.689 & 48.776 & 11.156 & 5.553\\
    \cmark & \cmark & \textbf{2.815} & \textbf{15.150} & \textbf{12.215} & \textbf{4.936} & \textbf{6.810} & \textbf{46.933} &  \textbf{11.744} & \textbf{5.973}\\
    \bottomrule
    \end{tabular}
    \caption{\textbf{Ablation of Sound Source Remixer conditions.} We achieve best performance with both CMSS and CLIP semantics.}
    \label{table:fold_ablation}
\end{table*}

\begin{table*}
    \fontsize{9pt}{9pt}\selectfont
    \centering
    \begin{tabular}{l | l l l l | l l l l}
    \toprule
    \multirow{2}{*}[-2pt]{$\alpha$} & \multicolumn{4}{c|}{Single-Source Generation} & \multicolumn{4}{c}{Multi-Source Generation} \\ \cmidrule(l{5pt}r{5pt}){2-5} \cmidrule(l{5pt}r{5pt}){6-9}
    & V-FAD$\downarrow$ & C-FAD$\downarrow$ & CS$\uparrow$ & SSMS$\uparrow$ & V-FAD$\downarrow$ & C-FAD$\downarrow$ & CS$\uparrow$ & SSMS$\uparrow$\\
    \midrule
    0 & 13.612 & 73.849 & 5.838 & 1.149 & 17.053 & 98.747 & 5.580 & 1.342\\
    0.35 & \textbf{2.815} & \textbf{15.150} & \textbf{12.215} & \textbf{4.936} & \textbf{6.810} & \textbf{46.933} &  \textbf{11.744} & \textbf{5.973}\\
    0.65 & 2.877 & 16.194 & 11.860 & 4.356 & 9.788 & 61.565 & 11.397 & 4.658\\
    1 & 3.323 & 16.740 & 11.299 & 4.075 & 10.098 & 60.810 & 11.585 & 4.237\\
    \bottomrule
    \end{tabular}
    \caption{\textbf{Ablation of CCMR.} We achieve the best performance with $\alpha=0.35$, which is used throughout other experiments.}
    \label{table:ccmr_ablation}
\end{table*}

\subsection{Ablation Study}
We conduct several ablation experiments to consolidate our claims in the Method section. We also provide an analysis on the learned CMSS manifold space and more ablations in Supplementary sections 3 and 4.

\smallskip
\noindent
\textbf{Effect of CMSS Manifold.} SS2A could learn to perform the V2A task without CMSS disambiguation. In order to prove the benefits of this disambiguation, we perturb the same Sound Source Remixer model with three different generation conditions: without CLIP embeddings, without CMSS embeddings, and with both embeddings. We train them on the same VGGSound data and evaluate the results with VGG-SS tests in \cref{table:fold_ablation}. A significant performance drop is observed in both generation fidelity and relevance when the CMSS conditioning is suppressed. This ablation confirms that CMSS disambiguation benefits our V2A task.

\smallskip
\noindent
\textbf{Effect of CCMR.} Recall that $\alpha$ controls CCMR's behavior in \cref{eq:ccmr}. When $\alpha=0$, the mask becomes identity and CCMR is stifled. We train the same CMSS manifold modules under four settings of $\alpha$ and conduct VGG-SS tests. \cref{table:ccmr_ablation} shows that with CCMR, we can enrich the CMSS semantics to benefit downstream generation. However, setting $\alpha$ to higher values degrades generation quality.

\subsection{Multimodal Sound Source Composition}

\begin{figure}
    \centering
    \includegraphics[width=\linewidth]{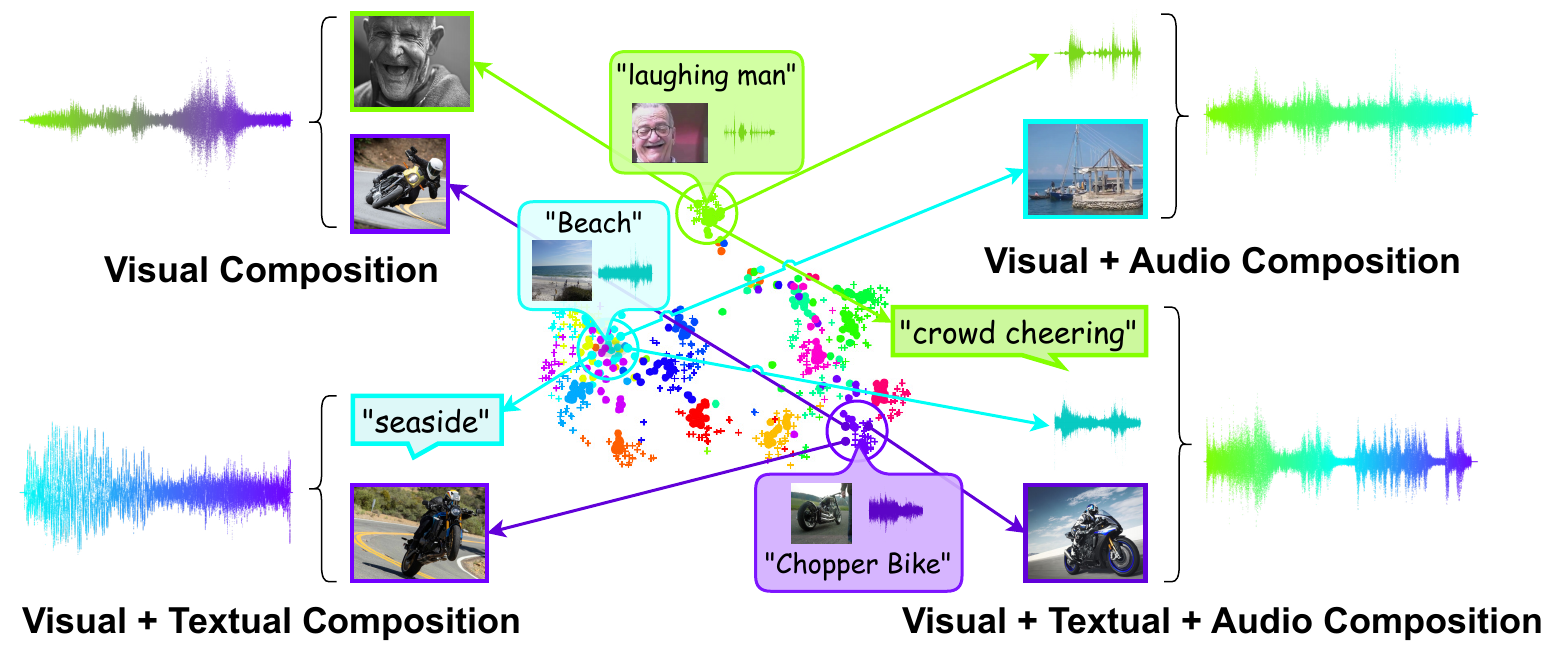}
    \caption{\textbf{Multimodal Sound Source Composition scenarios.} Our method can flexibly composite sound sources across visual, text, and audio modalities to guide V2A generation.}
    \label{fig:compose_scene}
\end{figure}

Since SS2A accepts sound source prompts as vision, text, and audio, we can intuitively control its generation by (1) editing specific sound sources and (2) compositing sources across modalities. We term this novel generation control scheme Multimodal Sound Source Composition. We show four visually-related composition scenarios in \cref{fig:compose_scene}. The composition results are best experienced via our website.

\smallskip
\noindent
\textbf{Visual Composition.} SS2A can generate realistic audio by composing visual sound sources. The result respects the supplied sources to render a convincing audio scene. For instance, we can synthesize a ``motorbike riders laughing" audio from pictures of a motorbike and a laughing man.

\smallskip
\noindent
\textbf{Visual-Text Composition.} SS2A can further control the V2A generation with textual semantics. For example, we can supply a ``motorbike" image and obtain a seaside riding audio with the text prompt ``seaside".

\smallskip
\noindent
\textbf{Visual-Audio Composition.} We can achieve a similar style control with audio semantics. For example, we can accompany a ``boat pier" image with a ``talking" audio to synthesize audio of a busy pier.

\smallskip
\noindent
\textbf{Visual-Text-Audio Composition.} We can synthesize audio with all three modalities involved. We have successfully produced a ``coastline motorcycle racing" audio with a motorcycle image, a ``crowd cheering" text, and a ``beach" audio.

\section{Conclusion}
In this work, we explore learning a sound source-aware audio generator, SS2A, that supports multimodal conditioning. By explicitly modeling the source disambiguation process with a contrastive cross-modal manifold on single-source visual-audio pairs, we are able to significantly boost our method's generation fidelity and relevance. Consequently, SS2A achieves state-of-the-art image-to-audio performance in both objective and subjective evaluations. With a simple temporal aggregation mechanism, SS2A also achieves competitive performance in video-to-audio synthesis. Moreover, we demonstrate the intuitive control of our generator in composition experiments of vision, text, and audio sound sources. During the learning of our manifold, we curate a new single-sound-source visual-audio dataset VGGS3. Additionally, we contribute a novel Sound Source Matching Score that measures  fine-grained audio-audio relevance with sound source detection. As SS2A is a fresh exploration, we discuss its limitations in Supplementary Section 8.

\clearpage
\twocolumn[
\begin{center}
\LARGE\bfseries Gotta Hear Them All: Towards Sound Source Aware Audio Generation \\
\vspace{2mm}
\Large Supplementary Materials
\vspace{1em}
\end{center}
]

\setcounter{secnumdepth}{2}

\section{Baseline Modifications} \label{sec:baseline_mod}
We choose Seeing and Hearing \cite{SandH} (S\&H)'s image-to-audio (I2A) branch as a baseline. However, we notice this branch also depends on text captioned from a large vision-language model, QWEN \cite{qwen}, on the input image. The text modality creates extra information in the I2A task, which is unfair for other methods since V2A-Mapper \cite{V2A-Mapper} and our SS2A can also utilize the captions to refine results. Therefore, we rename the unfair version of S\&H as S\&H-Text, and suppress the QWEN captions to generate the fair set of baseline results, which is named S\&H in experiments.

We directly generate results from Im2Wav \cite{Im2Wav} as it is focused on the image-to-audio task only. We also leave the setup of V2A-Mapper unchanged. Additionally, we obtain oracle generation results in the VGG-SS \cite{VGG-SS} and MUSIC \cite{MUSIC} tests by passing the ground-truth audio clips through CLAP \cite{CLAP} and then AudioLDM \cite{AudioLDM}. We name this baseline Oracle. Aside from the ground-truth audio, the Oracle results can be regarded as generated from an audio synthesis model that exhausts AudioLDM’s potential for audio synthesis. We expect any method utilizing AudioLDM for downstream generation, \textit{i.e.}, our SS2A and V2A-Mapper, to be inferior in performance against Oracle.

\section{Model Training and Architectures} \label{sec:model_arch}

\begin{figure*}
    \centering
    \includegraphics[width=\linewidth]{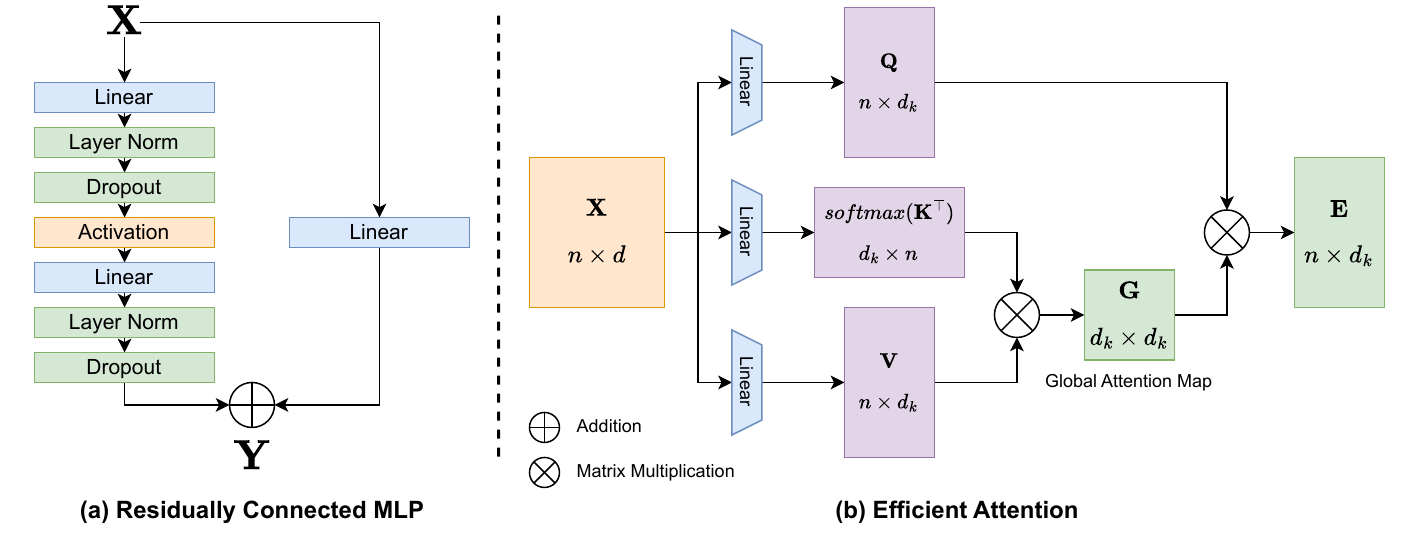}
    \caption{\textbf{Architecture of key module components.} We show a single instance instead of batch inference in (b).}
    \label{fig:arch}
\end{figure*}

We conduct all training on a Linux machine with a single RTX4090 GPU. Each module is trained to convergence, which takes 168 epochs for the CMSS manifold modules, 64 epochs for the Remixer module, and 128 epochs for the Temporal Aggregation module.

\subsection{Implementation Details}
We adopt the pretrained ViT-L/14 \cite{clip-git} for CLIP and the pretrained weights of audioldm-s-v2-full \cite{AudioLDM} for CLAP and AudioLDM. An open-source DALL$\cdot$E-2 Prior model \cite{dalle2-git} trained on the Aesthetics \cite{aesthetics} dataset translates the text-audio pairs. The visual detector is a YOLOv8x \cite{yolov8} model trained on the OpenImagesV7 \cite{OpenImagesV7} dataset with a 0.25 confidence threshold. We train all SS2A modules with an AdamW optimizer of 1e-4 learning rate until convergence and fix the classifier-free guidance's dropout rate to be 0.2.

\subsection{Cross-Modal Sound Source Manifold}

\paragraph{Architecture.}
We employ residually connected MLPs for the Cross-Modal Sound Source (CMSS) projectors and reconstructor, as shown in \cref{fig:arch} (a). We choose the ELU function for activations and the dropout probability as 0.2. To implement the reparameterization trick, we append two respective linear layers at each module's head to infer the estimated mean and variance. The output CMSS embeddings are sampled from a multivariate normal distribution with respect to these estimated parameters. The CMSS manifold's semantic dimension is fixed to be 768. The CLIP-ViT-L/14 dimension is 768 and the CLAP dimension is 512. The neuron numbers for each module's linear layers are reported in \cref{table:mlp_layers}. We conduct ablation experiments in \cref{subsec:fold_ablation} to obtain this optimal setup.

\begin{table}
    \centering
    \begin{tabular}{l l}
        \toprule
         Module & Linear Layer Neurons\\
         \midrule
         CLIP Projector & 768$\times$2, 1536$\times$2, 3072$\times$2\\
         CLAP Projector & 768$\times$2, 1536$\times$2, 3072$\times$2\\
         Reconstructor & 768$\times$2, 896$\times$2, 1024$\times$2, 2048$\times$2\\
         \bottomrule
    \end{tabular}
    \caption{\textbf{Neuron numbers for each CMSS module.} Note that we add a residual connection every two layers.}
    \label{table:mlp_layers}
\end{table}

\paragraph{Data Curation.}
We filter source-unannotated visual-audio pairs from the training set of VGGSound \cite{VGGS} with an open-vocabulary object segmentor, CLIP as RNN (CaR) \cite{CaR}, keeping the pairs where only one visual region is segmented. The confidence threshold of CaR is set to 0.5. We use the VGGSound category labels as segmentation vocabulary. CaR's pixel-level segmentations are abstracted into bounding boxes to capture fuller visual content. We crop each video's central frame with the predicted bounding box to pair with its audio clip and verify the data quality by manually reviewing 10 results from each category. The resulted VGGS3 dataset has 106514 samples across 221 sound source categories, which promises audio diversity. One potential bias is that the curated sound sources have unbalanced category frequencies with a max of 946 and a min of 100. We intend to release a balanced version of VGGS3 alongside the current version.

Additionally, we regard the SFX text-audio pairs from FSD50K \cite{FSD50K},  Epidemic Sound Effects \cite{Epidemic}, and BBC Sound Effects \cite{BBC} in the LAION-630K \cite{LAION630K} dataset as single-source since they have succinct label-like text captions. We translate their CLIP text embeddings to CLIP image space with the DALL$\cdot$E-2 Prior \cite{DALLE2} model to pair with their CLAP audio embeddings.

\paragraph{Mean-Teacher Training.}
The only manually-annotated single-source visual-audio pairs for our learning purpose are from the VGG Sound Source (VGG-SS) \cite{VGG-SS} dataset. The curated and translated pairs we collect can be regarded as noisy. We follow a Mean-Teacher \cite{MeanTeacher} paradigm in training for extra robustness. A teacher model is overfitted on the VGG-SS pairs to supervise another student model which sees the augmented/pseudo pairs during training. The teacher weights are updated by an exponential mean average schedule from student weights at each batch. We further filter out curated/translated pairs regarded as extremely noisy from student training by computing the cosine similarity between each pair's visual-audio CMSS embeddings with the teacher model and discarding the low-similarity ones adaptively with an elbow-finding algorithm Kneedle \cite{Kneedle}.

\subsection{Sound Source Remixer}

\paragraph{Efficient Attention.}
We adopt the Efficient Attention \cite{EfficientAtt} architecture in place of classical attention in the Sound Source Remixer and Temporal Aggreagation modules, which is shown in \cref{fig:arch} (b).  Instead of multiplying the query $\mathbf{Q}$ and key $\mathbf{K}^\top$ together for pairwise attention, the Efficient Attention computes a global attention map with value $\mathbf{V}$ as $softmax(\mathbf{K}^\top)\mathbf{V}$. The global attention map emphasizes the global context of tokens, which is desired since we already have rich individual audio semantics and only intend to mix them globally.

\paragraph{Architecture.}
Recall Eq. (8). We assign a learned [cls] token at the head of each token sequence for the Sound Source Remixer's prediction. The tokens first travel through a stack of attention modules, where each module contains an Efficient Attention layer followed by a feed forward network and an ELU activation. The [cls] token is then passed to two MLP heads to respectively estimate the mean and variance of the mixed CLAP audio embeddings. We then sample these embeddings from a normal distribution with these estimated parameters. The embeddings are further normalized with respect to their $l2$-norms to respect the original representation format of CLAP. Each MLP head is three-layer with [768, 640, 512] neurons and ELU activations. We use only one Efficient Attention layer following the optimal setup from ablation experiments in \cref{subsec:remixer_ablation}.

\paragraph{Temporal Aggregation Architecture.}
The Temporal Aggregation (TA) module employs the same optimal architecture setup as the Sound Source Remixer. We use the following formula to compute the positional embeddings:

\begin{algorithm}
\small
\caption{Cycle Mix} \label{alg:cyclemix}
\begin{algorithmic}
\Require $\left\{\mathbf{e}_m\right\}$, $\left\{\mathbf{x}_m\right\}$ \Comment{CMSS embs. and Remixer tokens}
\Require $T$ \Comment{user specified iterations}
\Require $N$ \Comment{user specified Remixer sample size}
\State $\mathbf{a}^\text{best}_\text{mix} \gets null$ \Comment{best Remixer generation}
\State $s \gets 0$ \Comment{best generation score}
\State $i \gets 0$
\While{$i < T$}
\State $\mathbf{a}_\text{src}^m \gets \chi\left(\mathbf{e}_m\right) \quad \forall \ m \in [1, \cdots, M]$
\State $\mathbf{a}_\text{mix}^n \gets sample\left[\psi\left(\mathbf{x}_1, \cdots, \mathbf{x}_{M+1}\right)\right] \quad \forall \ n \in [1, \cdots, N] $
\State $\mathbf{d}_n \gets \frac{1}{M}\sum sim(\mathbf{a}_\text{mix}^n, \mathbf{a}^m_\text{src}) \quad \forall \ n \in [1, \cdots, N]$\
\If{$\max \left(\mathbf{d}\right) > s$}
\State $\mathbf{a}^\text{best}_\text{mix} \gets \mathbf{a}_\text{mix}^{\arg \max \left(\mathbf{d}\right)}$
\State $s \gets r$
\State $\mathbf{x}_{M+1} \gets concat\left[\phi\left(\mathbf{a}^\text{best}_\text{mix}\right), \mathbf{0}\right]$ \Comment{conditions next iter.}
\EndIf
\State $i \gets i + 1$
\EndWhile
\State \Return $\mathbf{a}^\text{best}_\text{mix}$
\end{algorithmic}
\end{algorithm}

\begin{figure*}
    \centering
    \begin{minipage}{0.26\textwidth}
        \centering
        \includegraphics[width=\textwidth]{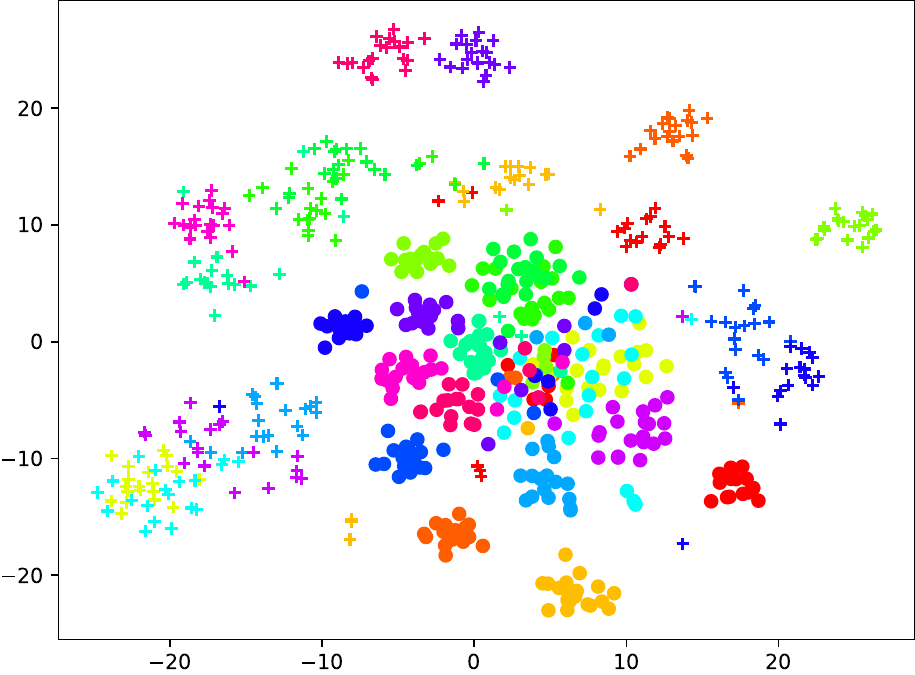}
        \caption*{Raw CLIP-CLAP Semantics}
    \end{minipage}
    \hfill
    \begin{minipage}{0.26\textwidth}
        \centering
        \includegraphics[width=\textwidth]{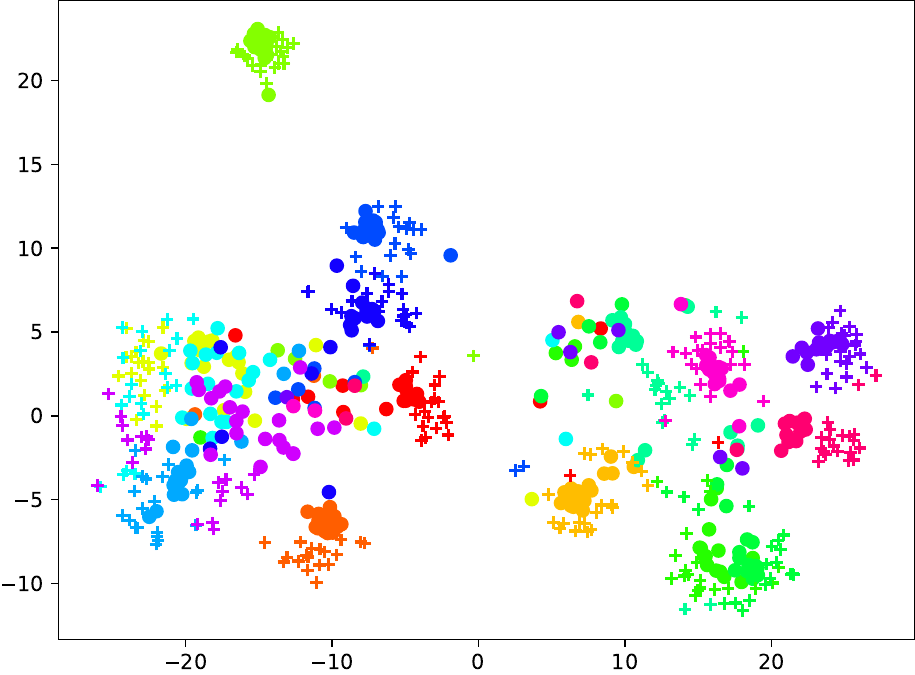}
        \caption*{CMSS Manifold Semantics}
    \end{minipage}
    \hfill
    \begin{minipage}{0.26\textwidth}
        \centering
        \includegraphics[width=\textwidth]{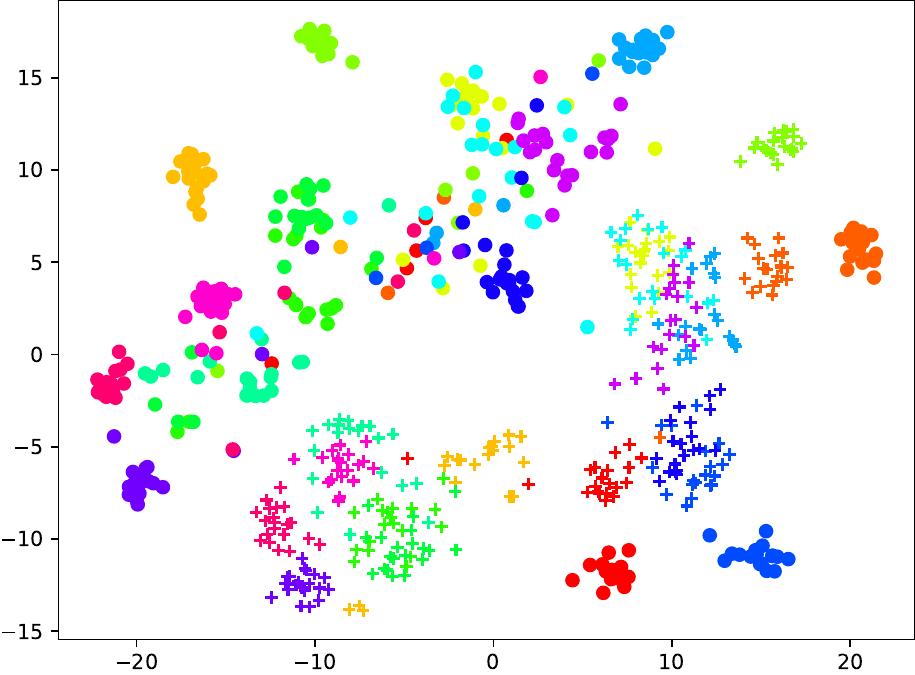}
        \caption*{Reconstructed CLAP Semantics}
    \end{minipage}
    \hfill
    \begin{minipage}{0.2\textwidth}
        \centering
        \includegraphics[width=\textwidth]{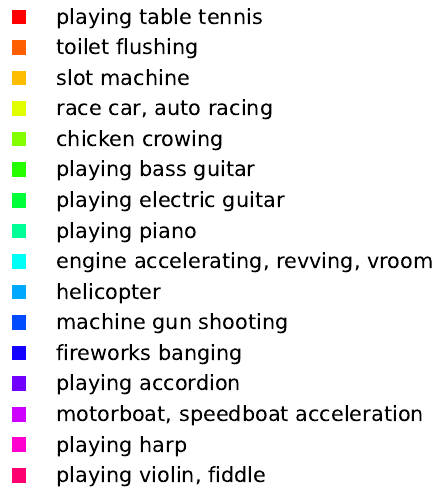}
    \end{minipage}
    \caption{\textbf{t-SNE visualizations of visual-audio modality alignment.} The first figure visualizes raw CLIP-CLAP embeddings, the second depicts their remapped CMSS manifold embeddings and the third illustrates reconstructed CLAP embeddings from CMSS manifold. The circles mark visual embeddings while the crosses mark audio embeddings.}
    \label{fig:fold_vis}
\end{figure*}

\paragraph{Cycle Mix Algorithm.}
Cycle Mix is an iterative mechanism that selects generated CMSS semantics to join the next mixing step. The scoring of these semantics is drawn from their cosine similarity with the previously mixed audio semantic, which can be illustrated as the algorithm below:

\begin{align}
    pos\left(2i, t\right) = \sin{\left(\frac{t}{1024^{2i/512}}\right)}, \\ pos\left(2i+1, t\right) = \cos{\left(\frac{t}{1024^{2i/512}}\right)},
\end{align}

\noindent where $i$ denotes the embedding position and $t$ is the integer timestamp of the video frame in $\left[1, 64\right]$. 1024 is fixed to be the positional embedding's frequency resolution, and 512 is the output CLAP embedding's dimension. To keep the model generative, we also model the TA module variationally with two prediction heads similar to those of the Sound Source Remixer.

\begin{figure*}
    \centering
    \includegraphics[width=\linewidth]{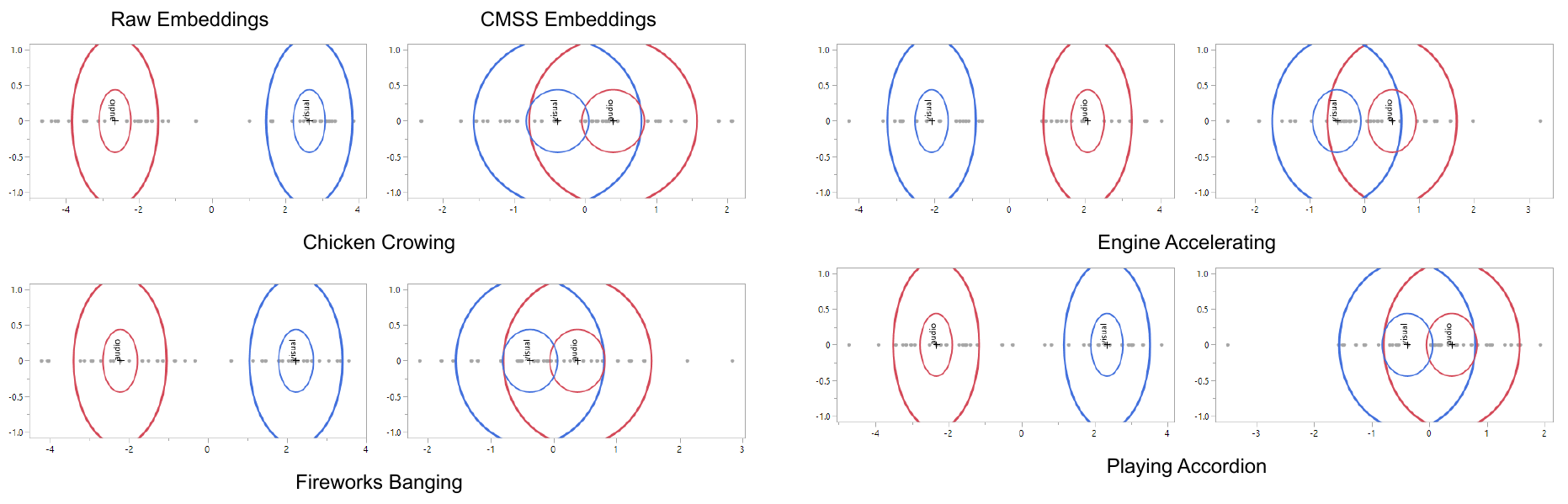}
    \caption{\textbf{Canonical plots of discriminant test.} The red inner circle marks the $95\%$ confidence interval and the red outer circle marks the $50\%$ normal contour of audio samples. The blue circles denote the visual samples.}
    \label{fig:discriminant}
\end{figure*}

\begin{table*}
    \centering
    \begin{tabular}{l | l | l l l}
        \toprule
        Category & Embedding & Percent Misclassified$\uparrow$ & Entropy $R^2 \downarrow$ & -2 Log-Likelihood$\uparrow$\\
        \midrule
        \multirow{2}{*}{Chicken Crowing} & Raw & 0.000 & 0.996 & 0.201\\
        & CMSS & \textbf{35.000} & \textbf{0.106} & \textbf{49.587}\\ \cmidrule(l{5pt}r{5pt}){1-1} \cmidrule(l{5pt}r{5pt}){2-2} \cmidrule(l{5pt}r{5pt}){3-5}
        \multirow{2}{*}{Engine Accelerating} & Raw & 0.000 & 0.985 & 0.846\\
        & CMSS & \textbf{32.500} & \textbf{0.100} & \textbf{49.884}\\ \cmidrule(l{5pt}r{5pt}){1-1} \cmidrule(l{5pt}r{5pt}){2-2} \cmidrule(l{5pt}r{5pt}){3-5}
        \multirow{2}{*}{Fireworks Banging} & Raw & 0.000 & 0.999 & 0.048\\
        & CMSS & \textbf{35.000} & \textbf{0.146} & \textbf{47.369}\\ \cmidrule(l{5pt}r{5pt}){1-1} \cmidrule(l{5pt}r{5pt}){2-2} \cmidrule(l{5pt}r{5pt}){3-5}
        \multirow{2}{*}{Playing Accordion} & Raw & 0.000 & 0.984 & 0.889\\
        & CMSS & \textbf{35.000} & \textbf{0.107} & \textbf{49.530}\\
        \bottomrule
    \end{tabular}
    \caption{\textbf{Statistics of discriminant test.} There are 20 visual samples and 20 audio samples in each category.}
    \label{table:discriminant}
\end{table*}

\begin{figure*}
    \centering
    \begin{minipage}{0.51\linewidth}
        \centering
        \includegraphics[width=\linewidth]{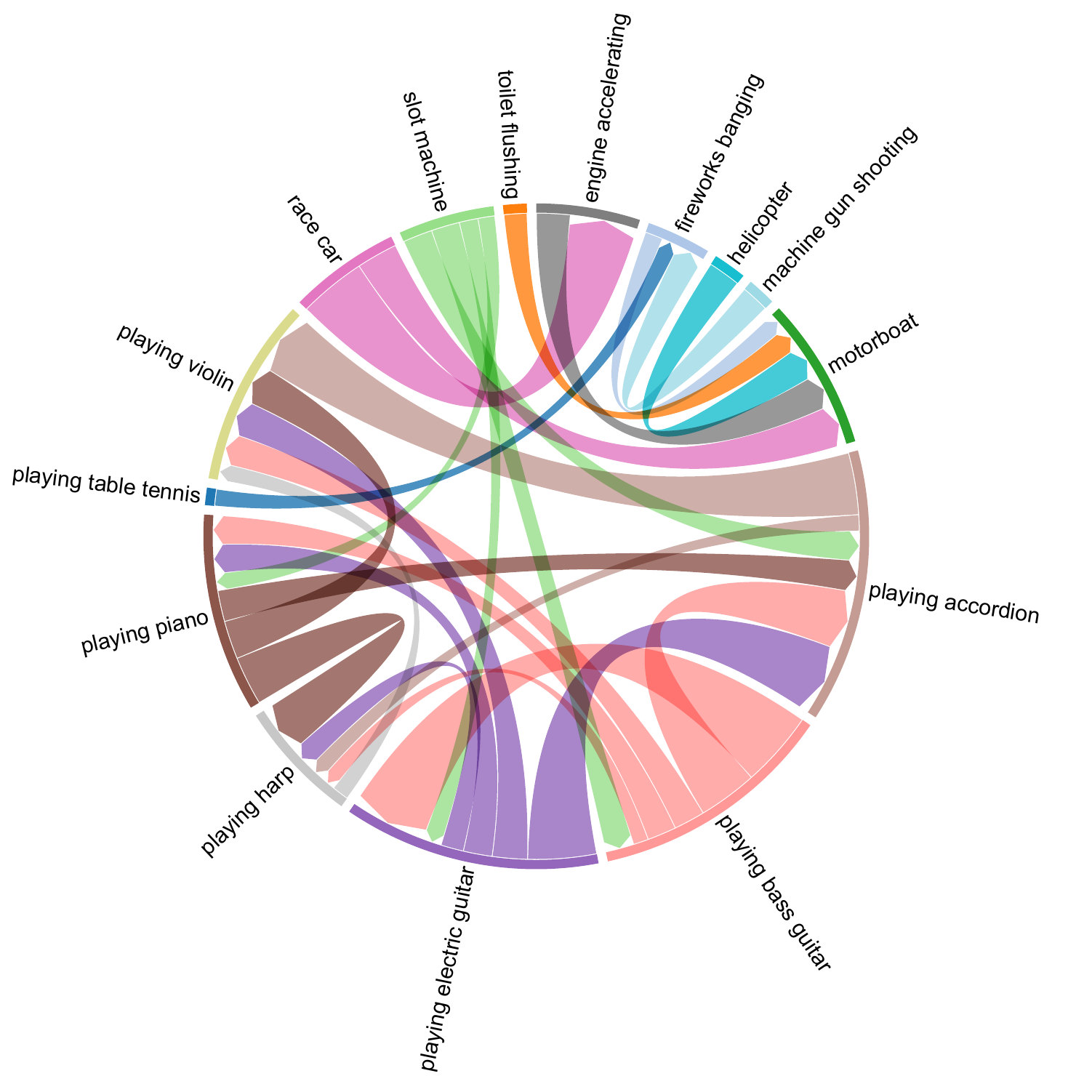}
        \caption{\textbf{Chord diagram of CMSS sound source similarities.} Wider chords indicate higher similarities between sources.}
        \label{fig:chord}
    \end{minipage}
    \begin{minipage}{0.48\linewidth}
        \centering
        \begin{tabular}{l l}
        \toprule
        Embeddings &  Partition Coefficient$\uparrow$\\
        \midrule
        Raw CLIP-CLAP & 8.473\\
        CMSS Manifold & 12.615\\
        Reconstructed CLAP & \textbf{12.800}\\
        \bottomrule
        \end{tabular}
        \captionof{table}{\textbf{Partition Coefficient test.} The reconstructed CLAP embeddings have the highest partition coefficient.}
        \label{table:partitionc}
    \end{minipage}
\end{figure*}

\section{Manifold Analysis} \label{sec:manifold_analysis}
We conduct a manifold analysis to better understand the behaviors of CMSS manifold (abbreviated as manifold below). Ideally, we would like to observe the following traits from this manifold: (1) modality gap between audio and visual sound sources is closed, and (2) clustering forms naturally for similar audio-visual sound sources. The first trait confirms the cross-modal alignment of CMSS embeddings. The second trait manifests the manifold's capability to disambiguate sound sources. To examine these effects, we randomly select 20 samples from each of the 16 top-occurring classes in the curated VGGS3 and report three experiments: visualizations, modality alignment tests, and clustering tests. We show that all three experiments support the existence of both traits in the manifold.

\subsection{Visualizations}

\paragraph{t-SNE Visualizations.}
Our visualizations are illustrated in \cref{fig:fold_vis}. To visualize the unprocessed CLIP and CLAP embeddings of sampled visual-audio pairs, we reduce the CLIP embeddings from 768 to 512 dims by Principal Component Analysis and visualize them together with the CLAP embeddings in t-SNE. We then respectively visualize the manifold embeddings of these samples and their reconstructed CLAP embeddings. It can be observed that modality gap is closed in CMSS manifold embeddings since the visual and audio embeddings are pulled towards each other. Furthermore, a natural clustering forms for each audio category in the manifold space.

Although both desired traits are still present in the reconstructed CLAP embeddings, we observe that the modality gap is larger and clustering is less prominent. Therefore, we choose to operate the Sound Source Remixer on manifold embeddings instead of the reconstructed CLAP embeddings for audio synthesis.

\paragraph{Sound Source Similarity Visualization.}

We assign each visual sample to a cluster based on its audio class label. Average linkages in terms of cosine similarity are computed between clusters. We then filter these linkages with a $>0.4$ threshold to visualize clusters that are very close to each other as a chord diagram in \cref{fig:chord}. We observe that our CMSS embeddings encode audio traits of sound sources as well as visual traits, which is the objective of our auxiliary reconstruction in CMSS manifold. For instance, although a machine gun is visually different from fireworks, our manifold picks up the information that they share audio similarity. Likewise, the musical instruments are more similar in manifold space than other sound sources.

\subsection{Modality Alignment Tests}

\paragraph{Discriminant Test.}
We randomly select 4 audio categories and conduct a discriminant test on each in \cref{table:discriminant}. These results are also visualized as a canonical plot in \cref{fig:discriminant}. The discriminant model is a wide linear binary classifier to predict whether a given sample is from the visual or audio modality. We observe that this classifier works perfectly on raw embedding samples but fails to classify the CMSS embeddings with a low Entropy $R^2$ (classification contingency) and high -2 log-likelihood (classification uncertainty). This discriminant test supports the manifold's ability to close the modality gap between visual and audio data distributions, as the discriminant classifier is significantly confused after the manifold remapping of CLIP and CLAP embeddings.

\subsection{Clustering Tests}

\paragraph{Partition Coefficient Test.}

To examine whether our manifold embeddings have a stronger clustering tendency than the raw CLIP-CLAP embeddings, we evaluate Partition Coefficient (PC) \cite{clustervalid} as a clustering validation index. Our PC is computed as:

\begin{equation}
    PC = \frac{1}{N}\sum^N_{i=1}\sum^{M}_{j=1}u^2_{ij},
\end{equation}

\noindent where $N$ is the sample size, $M$ is the number of clusters and $u_{ij}$ is the membership value of sample $i$ to cluster $j$. Unlike the classic situation, where clusters are not assigned, we do have this information beforehand as the samples' audio class labels. As such, we find the centroid of each cluster by taking the average of its samples, and define $u$ as the cosine similarity between each sample and each centroid. Moreover, since cosine similarity can have negative values whose squaring confuses PC, we linearly rescale the cosine similarity from $\left[-1, 1\right]$ to $\left[0, 1\right]$. We find that both manifold embeddings and the reconstructed CLAP embeddings obtain significantly higher PCs than the raw embeddings, as recorded in \cref{table:partitionc}. This evaluation supports our claim that the manifold processing enhances clustering of sound source semantics.

\section{More Ablations} \label{sec:more_ablations}

\begin{table*}
    \centering
    \begin{tabular}{c c | l l l l | l l l l}
    \toprule
    \multirow{2}{*}[-2pt]{Sample Size} & \multirow{2}{*}[-2pt]{Iterations} & \multicolumn{4}{c|}{Single-Source Generation} & \multicolumn{4}{c}{Multi-Source Generation} \\ \cmidrule(l{5pt}r{5pt}){3-6} \cmidrule(l{5pt}r{5pt}){7-10}
    & & V-FAD$\downarrow$ & C-FAD$\downarrow$ & CS$\uparrow$ & MS$\uparrow$ & V-FAD$\downarrow$ & C-FAD$\downarrow$ & CS$\uparrow$ & MS$\uparrow$\\
    \midrule
    1 & 1 & \textbf{2.683} & \underline{14.869} & 12.364 & \textbf{5.048} & 6.864 & 47.219 & \textbf{11.945} & 5.868\\
    4 & 4 & 2.762 & \textbf{14.687} & 12.356 & 4.921 & \textbf{6.476} & 48.758 & 11.721 & 5.868\\
    4 & 64 & 2.713 & 14.980 & 12.377 & \underline{5.042} & 6.814 & 48.789 & \underline{11.912} & \underline{5.921}\\
    64 & 4 & \underline{2.710} & 15.069 & \underline{12.390} & 5.037 & 6.813 & \underline{46.963} & 11.907 & \underline{5.921}\\
    64 & 64 & 2.815 & 15.150 & 12.215 & 4.936 & 6.810 & \textbf{46.933} & 11.744 & \textbf{5.973}\\
    64 & 256 & 2.720 & 15.567 & 12.190 & 4.912 & \underline{6.755} & 47.657 & 11.708 & 5.816\\
    256 & 64 & 2.762 & 15.098 & \textbf{12.404} & 4.943 & 6.890 & 49.605 & 11.624 & 5.789\\
    256 & 256 & 2.876 & 15.728 & 12.060 & 5.022 & 7.220 & 47.373 & 11.577 & 5.684\\
    \bottomrule
    \end{tabular}
    \caption{\textbf{Ablation of Cycle Mix.} We choose both sample size and iterations to be 64 as the optimal parameter setup.}
    \label{table:cycle_ablation}
\end{table*}

\begin{table*}
    \centering
    \begin{tabular}{c | l l l l | l l l l}
    \toprule
    CMSS Variant & \multicolumn{4}{c|}{Projector Layers} & \multicolumn{4}{c}{Reconstructor Layers}\\
    \midrule
    A & \multicolumn{4}{l|}{768$\times$2, 1536$\times$2} & \multicolumn{4}{l}{768$\times$2, 896$\times$2, 1024$\times$2}\\
    B & \multicolumn{4}{l|}{768$\times$2, 1536$\times$2, 3072$\times$2} & \multicolumn{4}{l}{768$\times$2, 896$\times$2, 1024$\times$2, 2048$\times$2}\\
    C & \multicolumn{4}{l|}{768$\times$2, 1536$\times$4, 3072$\times$2} & \multicolumn{4}{l}{768$\times$2, 896$\times$3, 1024$\times$3, 2048$\times$2}\\
    \midrule
    \multirow{2}{*}[-2pt]{CMSS Variant} & \multicolumn{4}{c|}{Single-Source Generation} & \multicolumn{4}{c}{Multi-Source Generation} \\ \cmidrule(l{5pt}r{5pt}){2-5} \cmidrule(l{5pt}r{5pt}){6-9}
    & V-FAD$\downarrow$ & C-FAD$\downarrow$ & CS$\uparrow$ & MS$\uparrow$ & V-FAD$\downarrow$ & C-FAD$\downarrow$ & CS$\uparrow$ & MS$\uparrow$\\
    \midrule
    A & 6.684 & 42.423 & 7.974 & 3.220 & 9.036 & 54.437 & 10.699 & 4.868\\
    B & \textbf{2.815} & \textbf{15.150} & \textbf{12.215} & \textbf{4.936} & \textbf{6.810} & \textbf{46.933} & \textbf{11.744} & \textbf{5.973}\\
    C & 6.945 & 26.035 & 9.868 & 3.993 & 11.343 & 59.819 & 9.908 & 4.763\\
    \bottomrule
    \end{tabular}
    \caption{\textbf{Ablation of CMSS architectures.} We choose model variant B as the optimal parameter setup.}
    \label{table:fold_arch_ablation}
\end{table*}

\begin{table*}
    \centering
    \begin{tabular}{c | l l l l | l l l l}
    \toprule
    \multirow{2}{*}[-2pt]{Attention Layers} & \multicolumn{4}{c|}{Single-Source Generation} & \multicolumn{4}{c}{Multi-Source Generation} \\ \cmidrule(l{5pt}r{5pt}){2-5} \cmidrule(l{5pt}r{5pt}){6-9}
    & V-FAD$\downarrow$ & C-FAD$\downarrow$ & CS$\uparrow$ & MS$\uparrow$ & V-FAD$\downarrow$ & C-FAD$\downarrow$ & CS$\uparrow$ & MS$\uparrow$\\
    \midrule
    1 & 2.815 & 15.150 & 12.215 & 4.936 & \textbf{6.810} & \textbf{46.933} & \textbf{11.744} & \textbf{5.973}\\
    2 & 2.685 & 14.839 & 12.545 & \textbf{5.116} & 8.634 & 52.697 & 11.583 & 5.105\\
    4 & \textbf{2.378} & \textbf{14.021} & \textbf{12.716} & 4.916 & 8.119 & 55.639 & 11.056 & 5.395\\
    \bottomrule
    \end{tabular}
    \caption{\textbf{Ablation of the Sound Source Remixer architecture.} We choose one attention layer as the optimal setup.}
    \label{table:remixer_ablation}
\end{table*}

\subsection{Ablation of Cycle Mix}
The Cycle Mix algorithm has two adjustable parameters: the number of iterations and the the sampling size of remixed CLAP embeddings in each iteration. We conduct 8 VGG-SS tests to observe the effect of Cycle Mix parameters. These ablations are illustrated in \cref{table:cycle_ablation}. When the iteration is 1 and sampling size is 1, we directly obtain the Remixer's output without Cycle Mix. Higher sample size and iterations lead to better multi-source generations while single-source performance slightly drops. Setting the parameters too high compromises performance in both generation tasks. Since this work's primary focus is multi-sound-source audio synthesis, we determine a sample size of 64 and an iteration count of 64 as the optimal parameter setup and use it throughout other experiments.

\subsection{Ablation of CMSS Architectures} \label{subsec:fold_ablation}
We find the optimal architectures of the CMSS manifold modules through ablation experiments illustrated in \cref{table:fold_arch_ablation}. These ablations are performed by training the SS2A pipeline with different CMSS module configurations and the same Sound Source Remixer. We observe that shallower projectors and reconstructor underfit on the training data while deeper modules tend to overfit. Consequently, we choose CMSS model variant B in \cref{table:fold_arch_ablation} to be our optimal setup and use it throughout other experiments.

\subsection{Ablation of Remixer Architecture} \label{subsec:remixer_ablation}
The optimal setup of the Sound Source Remixer's architecture is found through ablations recorded in \cref{table:remixer_ablation}. Increasing the number of attention layers slightly increases single-source generation performance. However, the multi-source generation quality is significantly sacrificed as attention layers stack deeper. Since our work's primary focus is to tackle the multi-sound-source audio generation problem, we choose the one-attention-layer architecture as the optimal setting for the Sound Source Remixer. We use the same architecture for the Temporal Aggregation module.

\section{More Experiments}

\begin{table*}
\centering
\fontsize{9pt}{9pt}\selectfont
\begin{tabular}{l | l | l l l l | l l l l}
    \toprule
      & \multirow{2}{*}[-2pt]{Method} & \multicolumn{4}{c|}{VGG-SS} & \multicolumn{4}{c}{MUSIC}\\ \cmidrule(l{5pt}r{5pt}){3-6} \cmidrule(l{5pt}r{5pt}){7-10}
      & & V-FAD$\downarrow$ & C-FAD$\downarrow$ &  CS$\uparrow$ & SSMS$\uparrow$ & V-FAD$\downarrow$ & C-FAD$\downarrow$ &  CS$\uparrow$ & SSMS$\uparrow$\\
     \midrule
     \multirow{6}{*}{\rotatebox{90}{Single-Source}} & Diff-Foley & 7.212 & 39.309 & 11.045 & 4.099 & 27.633 & 79.068 & 9.286 & 5.899\\
     & Frieren & 2.799 & 28.340 & 11.134 & 4.326 & 9.001 & 56.014 & 12.613 & 6.650\\
     & MultiFoley & 3.890 & 30.244 & 11.827 & \underline{5.648} & - & - & - & -\\
     & MMAudio & \underline{2.882} & \underline{20.659} & \textbf{13.971} & \textbf{5.991} & \textbf{3.471} & \underline{33.339} & \textbf{14.662} & \textbf{7.866}\\ \cmidrule(l{5pt}r{5pt}){2-2} \cmidrule(l{5pt}r{5pt}){3-6} \cmidrule(l{5pt}r{5pt}){7-10}
     & SS2A (Ours) & \textbf{2.270} & \textbf{14.433} & \underline{12.176} & 5.336 & \underline{8.353} & \textbf{25.393} & \underline{13.392} & \underline{7.182}\\ \cmidrule(l{5pt}r{5pt}){1-2} \cmidrule(l{5pt}r{5pt}){3-6} \cmidrule(l{5pt}r{5pt}){7-10}
     \multirow{6}{*}{\rotatebox{90}{Multi-Source}} & Diff-Foley & 13.373 & 75.829 & 12.209 & 4.789 & 12.423 & 105.299 & 8.561 & 4.843\\
     & Frieren & 10.601 & 66.896 & 11.884 & 4.526 & \underline{4.803} & 69.921 & 11.857 & 5.979\\
     & MultiFoley & 9.082 & 67.201 & \underline{12.648} & 5.621 & - & - & - & -\\
     & MMAudio & \underline{8.178} & \underline{50.657} & \textbf{13.425} & \textbf{6.053} & 5.236 & \underline{31.880} & \textbf{13.149} & \textbf{6.693}\\ \cmidrule(l{5pt}r{5pt}){2-2} \cmidrule(l{5pt}r{5pt}){3-6} \cmidrule(l{5pt}r{5pt}){7-10}
     & SS2A (Ours) & \textbf{6.837} & \textbf{45.647} & 12.195 & \underline{5.658} & \textbf{3.562} & \textbf{30.260} & \underline{12.436} & \underline{6.021}\\
     \bottomrule
\end{tabular}
\caption{\textbf{Video to audio comparisons.} The first and second places are \textbf{bolded} and \underline{underlined}, respectively.}
\label{table:video-objective}
\end{table*}

\begin{table}
\centering
\begin{tabular}{l | c c}
    \toprule
    \multirow{2}{*}[-2pt]{Method} & \multicolumn{2}{c}{WMAO $\downarrow$}\\
    \cmidrule(l{5pt}r{5pt}){2-3}
    & Top-1 & Top-5\\
    \midrule
    Diff-Foley & \textbf{1.227} & \textbf{1.127}\\
    FoleyCrafter & \textit{1.247} & \underline{1.167}\\
    V2A-Mapper & 1.299 & 1.197\\
    \cmidrule(l{5pt}r{5pt}){1-1} \cmidrule(l{5pt}r{5pt}){2-3}
    Ours w/o TA & 1.292 & 1.205\\
    Ours & \underline{1.243} & \textit{1.172}\\
    \bottomrule
\end{tabular}
\captionof{table}{\textbf{AVSync synchronization tests.} \textbf{Bold}, \underline{underline}, and \textit{italic} mark first, second, and third placements.}
\label{table:offset}
\end{table}

\subsection{Video-to-Audio Objective Tests}
We employ the same metrics in image-to-audio tasks to judge the generation quality of video-to-audio samples. Our method performs competitively in \cref{table:video-objective}, especially in multi-source scenarios. MultiFoley isn’t open-sourced, so MUSIC tests are void. We lead in fidelity while MMAudio leads in relevance, likely due to its joint audio-visual (AV) and audio-text (AT) training. Our Remixer, trained only on AV, aligns more with AV distributions (fidelity), while MMAudio benefits from semantics in AT (relevance).

\subsection{Video-to-Audio Synchronization Tests}

\paragraph{Weighted Mean Absolute Offset.}

To prove the efficacy of our TA mechanism for video-audio synchronization, we employ SynchFormer \cite{SyncFormer} to predict the temporal Weighted Mean Abolute Offset (WMAO) in seconds between the original video and the generated audio on the AVSync15 \cite{AVSync15} dataset. We weight and sum the top-k predictions of SyncFormer with their confidence scores. A lower WMAO perceives smaller drifts between video-audio signals, indicating higher synchronization.

Our method generates competitive video-audio synchronization as shown in \cref{table:offset}. The ablation in TA mechanism shows that it is key to SS2A's temporal alignment capability. Moreover, the results indicate that our nonlinear TA function performs better than V2A-Mapper's linear setup.

\section{Setup of Subjective Evaluation} \label{sec:qual_details}

\begin{figure*}
    \centering
    \includegraphics[width=\linewidth]{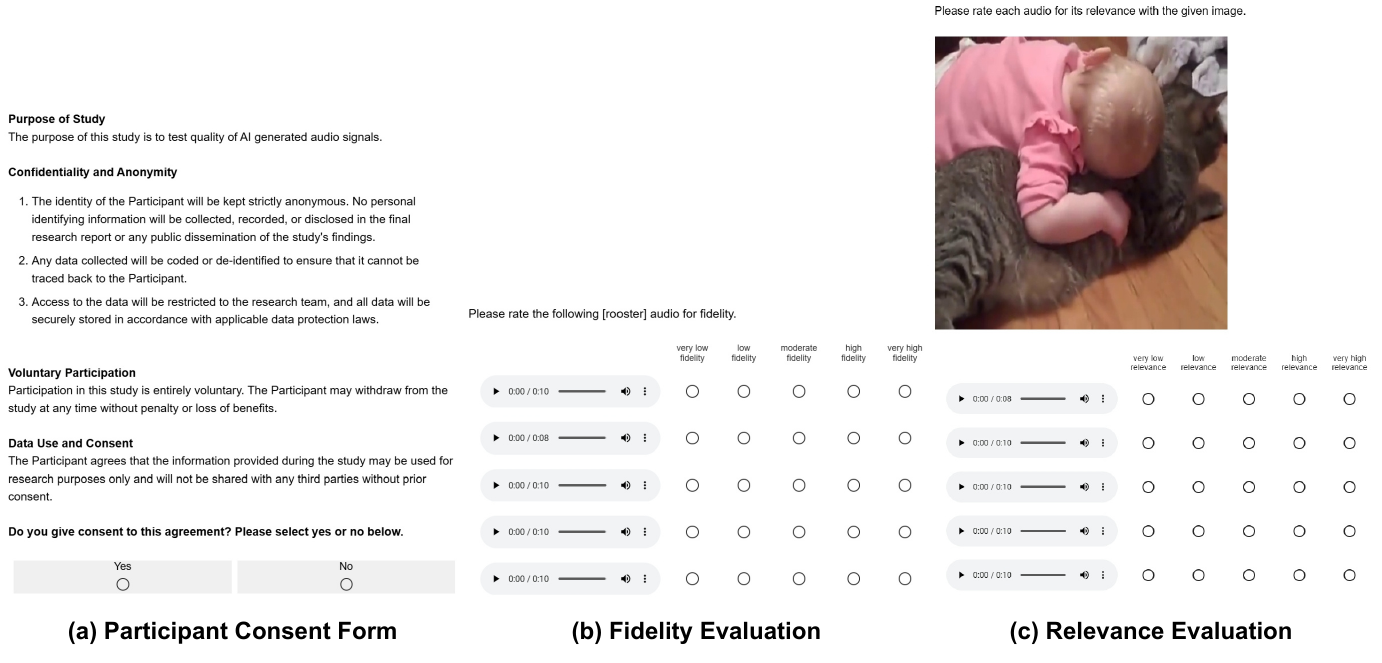}
    \caption{\textbf{Screenshots of subjective survey.} Each row of circles prompts a single-choice question to the testee.}
    \label{fig:survey}
\end{figure*}

We disseminate an online survey for the subjective evaluation and collect results from 20 participants to measure generation fidelity and relevance of our method along with baselines. The baseline methods include Im2Wav, S\&H-Text, and V2A-Mapper as described in \cref{sec:baseline_mod}. In the first survey section, we ask the participants to sign a consent form as illustrated in \cref{fig:survey} (a). The non-consenting participants are screened out without any data collection. In the second section, we ask 20 fidelity-rating questions without visual cues following \cref{fig:survey} (b). To unify the comparison context, we include a short tag in the question describing the ground-truth audio content. For each generated sample, the testee is asked to give out fidelity rating on a 1-5 scale. In the third section, we ask 20 relevance-rating questions given the visual condition used during generation, which is depicted in \cref{fig:survey} (c). The ratings are also collected on a 1-5 scale.

After data collection, we thoroughly anonymize the participant information to deidentify any personal data. We then compute the Mean Opinion Score (MOS) \cite{MOS} respectively from the fidelity and relevance ratings.


\section{Why Not Cascaded Composition?}
The most straightforward way to composite multimodal sound sources into a single audio is to generate an audio track for each source condition via video-to-audio or text-to-audio models and overlay them together. However, such a cascading audio synthesis system lacks interaction, context and style awareness when integrating multiple sound sources, which are keys to a convincing audio scene. We show with qualitative examples in this section that our SS2A composition achieves these features. These examples are best experienced on our website at the Composition Comparisons section.

\begin{figure}
    \centering
    \includegraphics[width=\linewidth]{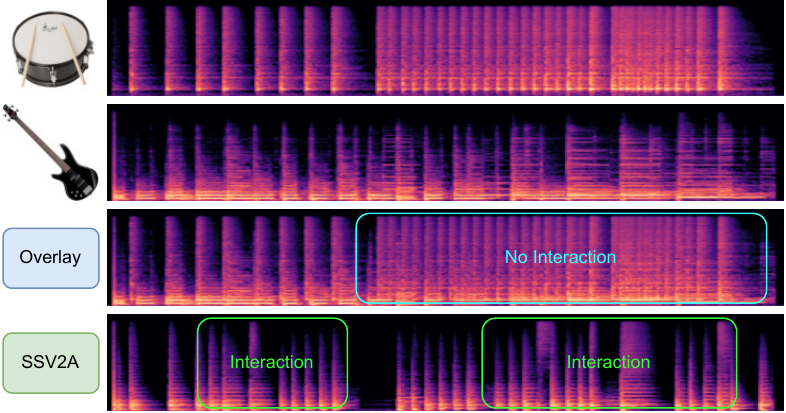}
    \caption{\textbf{Example of interaction awareness.} Our composition arranges drum and bass sounds into interactive music.}
    \label{fig:interact}
\end{figure}

\begin{figure}
    \centering
    \includegraphics[width=\linewidth]{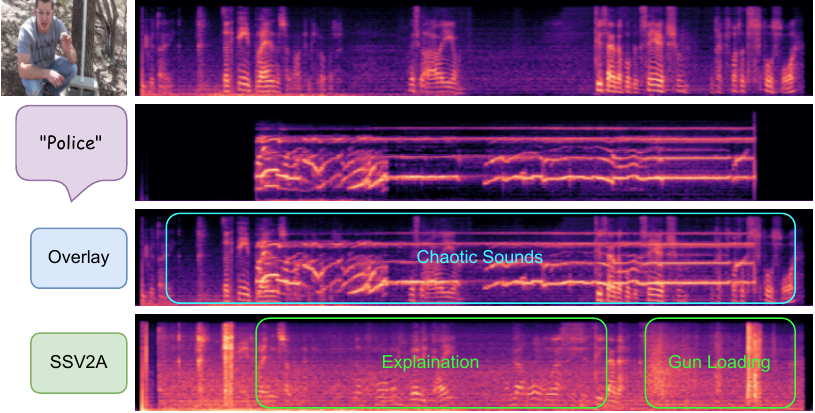}
    \caption{\textbf{Example of context awareness.} Our composition transforms a normal talking man into a police officer on duty.}
    \label{fig:context}
\end{figure}

\begin{figure}
    \centering
    \includegraphics[width=\linewidth]{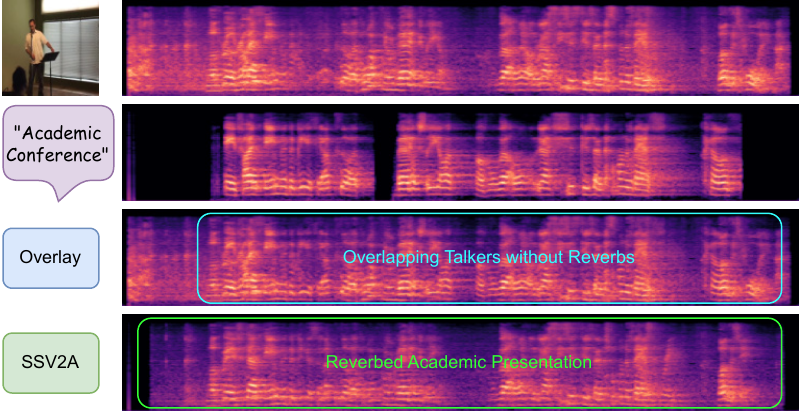}
    \caption{\textbf{Example of style awareness.}  Our composition changes a normal speech into an academic presentation with conference room reverb.}
    \label{fig:style}
\end{figure}

\subsection{Interaction Awareness}
We generate a drum-only audio clip and a bass-only clip with SS2A. Simply overlaying them yields a mixed track with no interactions between these instruments, as shown in \cref{fig:interact}. With SS2A composition, we are able to generate an appealing piece of drum-bass music with rich interactions.

\subsection{Context Awareness}
We generate audio clips of a normal speech and an academic conference in \cref{fig:context}. Cascaded composition synthesizes an audio clip with conflicting talkers and no conference room reverb. Our SS2A composition properly transfers the speech style into a reverberated academic presentation.

\subsection{Style Awareness}
In \cref{fig:style}, we generate an audio clip of a talking man and another clip of police activities. The cascaded composition yields chaotic sounds as the police event context is not perceived. Our SS2A composition successfully picks up this global cue and generates a police officer's voice followed by a gun loading/shooting sound to indicate police events.

\section{Limitations and Future Improvements}
Two limitations exist in SS2A. First, we address the video-to-audio synchronization in SS2A with a naive temporal module. Existing works \cite{SparseSync, SyncFormer} show that temporal alignment is a nontrivial problem due to the sparsity of synchronization cues in both time and space. Second, we observe that SS2A is less sensitive to audio conditioning than visual or text inputs. We suspect that this phenomenon is due to the lack of CLIP semantics when the Sound Source Remixer is prompted with audio conditions. We propose solutions to these issues as future works in this section.

\subsection{Temporal Synchronization}
Our simple TA module has shown decent synchronization capability by attending to global audio scene semantics, as evidenced by WMAO evaluations in the Tab. 3. In reality, each sound source can have different temporal ``activation" intervals in an audio scene. For example, a dog may bark in only the first three seconds of a video, followed by a baby laughing. In future works, we propose to perceive each sound source's activations locally in time, similar to the tracklet detection \cite{tracklet} concept in video multi-object tracking. With this finer-grained temporal aggregation approach, we aim to further enhance SS2A's temporal alignment performance.

\subsection{Sensitivity of Audio Conditioning}
As stated in the limitations section, the lower sensitivity of our method's audio conditioning compared to other modalities' input is due to the lack of CLIP semantics for audio sound source inputs. An existing method, Wav2CLIP \cite{Wav2CLIP}, translates CLAP audio embeddings to relevant CLIP embeddings. However, it operates on ViT-B/32 instead of ViT-L/14, which is the CLIP variant SS2A employs. In future works, we plan to train a Wav2CLIP model compatible with SS2A to address the audio sensitivity issue.

\section{Ethical Statement}
Our human evaluation is strictly anonymized without collecting any sensitive personal data. We also obtain explicit verbal consent from participants by asking them to sign a data collection agreement form before survey and screening out non-consenting participants. We intend to make our curated dataset, VGGS3, publicly available to contribute to the visual-audio research community. Our method complements videos and images with convincing audio tracks. Its application may have malicious outcomes in deepfake multimedia products if used without censorship. Multiple multimedia deepfake detection approaches have been proposed including audio deepfake detection \cite{audiofake}. We are committed to contribute generation samples for strengthening the learning of these detectors.

\bibliography{aaai2026}

\end{document}